\documentclass[sigconf,balance=false]{acmart}

\renewcommand\footnotetextcopyrightpermission[1]{} 
\usepackage{hyperref}
\usepackage{cleveref}
\usepackage{multirow}
\usepackage[inline]{enumitem}
\usepackage{subcaption}
\usepackage[inline]{enumitem}
\usepackage{array}

\begin{document}

\title{
Poisoning Network Flow Classifiers
}

\author{Giorgio Severi$^*$,\hspace{1em} Simona Boboila$^*$,\hspace{1em} Alina Oprea$^*$,\vspace{0.3em}\\ \hspace{1em} John Holodnak$^{\mathsection}$,\hspace{1em} Kendra Kratkiewicz$^{\mathsection}$,\hspace{1em} Jason Matterer$^{\mathsection}$ }
\affiliation{%
\vspace{0.5em}
\institution{Northeastern University$^*$ \hspace{0.3em} MIT Lincoln Laboratory$^\mathsection$}
\city{}
\state{}
\country{}
}

\begin{abstract}
As machine learning (ML) classifiers increasingly oversee the automated monitoring of network traffic, studying their resilience against adversarial attacks becomes critical. This paper focuses on poisoning attacks, specifically backdoor attacks, against network traffic flow classifiers. We investigate the challenging scenario of clean-label poisoning where the adversary's capabilities are constrained to tampering only with the training data --- without the ability to arbitrarily modify the training labels or any other component of the training process.  We describe a trigger crafting strategy that leverages model interpretability techniques to generate trigger patterns that are effective even at very low poisoning rates. 
Finally, we design novel strategies to generate stealthy triggers, including an approach based on generative Bayesian network models, with the goal of minimizing the conspicuousness of the trigger, and thus making detection of an ongoing poisoning campaign more challenging.
Our findings provide significant insights into the feasibility of poisoning attacks on network traffic classifiers used in multiple scenarios, including detecting malicious communication and application classification.
\end{abstract}

\settopmatter{printfolios=true} 

\maketitle
\pagestyle{plain}

\section{Introduction}

Automated monitoring of network traffic plays a critical role in the security posture of many companies and institutions. 
The large volumes of data involved, and the necessity for rapid decision-making, have led to solutions that increasingly rely on machine learning (ML) classifiers to provide timely warnings of potentially malicious behaviors on the network.
Given the relevance of this task, undiminished despite being studied for quite a long time~\cite{mukherjeeNetworkIntrusionDetection1994a}, a number of machine learning based systems have been proposed in recent years~\cite{mirskyKitsuneEnsembleAutoencoders2018a, ongunDesigningMachineLearning2019, ongunPORTFILERPortLevelNetwork2021, yangFeatureExtractionNovelty2021, hollandReproducibleNetworkTraffic2022} to classify network traffic.

The same conditions that spurred the development of new automated network traffic analysis systems, have also led researchers to develop adversarial machine learning attacks against them, targeting both deployed models~\cite{Handley2001,Cao2017,Ayub2020,papadopoulosLaunchingAdversarialAttacks2021,FENCE} (\emph{evasion} attacks) 
and, albeit to a lesser extent, their training process~\cite{apruzzeseAddressingAdversarialAttacks2019, liChronicPoisoningMachine2018a, ningTrojanFlowNeuralBackdoor2022, holodnakBackdoorPoisoningEncrypted2022} (\emph{poisoning} attacks).
We believe this second category is particularly interesting, both from an academic perspective as well as a practical one. 
Recent research on perceived security risks of companies deploying machine learning models repeatedly highlighted poisoning attacks as a critical threat to operational ML systems~\cite{sivakumarAdversarialMachineLearningIndustry2020, grosseMachineLearningSecurity2023}.
Yet, much of the prior research on poisoning attacks in this domain tends to adopt threat models primarily formulated in the sphere of image classification, such as assuming that the victim would accept a pre-trained model from a third party~\cite{ningTrojanFlowNeuralBackdoor2022}, thus allowing adversarial control over the entire training phase, or granting the adversary the ability to tamper with the training labels~\cite{apruzzeseAddressingAdversarialAttacks2019}.
As awareness of poisoning attacks permeates more extensively, it is reasonable to assume that companies developing this type of systems will exhibit an increased wariness to trust third parties providing pre-trained classifiers, and will likely spend resources and effort to control or vet both code and infrastructure used during training.
For this reason, we believe it is particularly interesting to focus on the less studied scenario of an adversary who is restricted to tampering only with the training data (\emph{data-only} attack) by disseminating a small quantity of maliciously crafted points, and without the ability to modify the labels assigned to training data (\emph{clean-label}) or any other component of the training process.

Our aim is to investigate the feasibility and effects of poisoning attacks, and in particular backdoor attacks ---where an association is induced between a trigger pattern and an adversarially chosen output of the model---, on network traffic flow classifiers.
Our approach focuses on the manipulation of aggregated traffic flow features rather than packet-level content, as they are common in traffic classification applications~\cite{mooreDiscriminatorsUseFlowbased2005,yangFeatureExtractionNovelty2021,ongunPORTFILERPortLevelNetwork2021}.
We will focus on systems that compute aggregated features starting from the outputs of the network monitoring tool Zeek\footnote{\url{https://zeek.org/} Previously known as Bro.}, because of its large user base.
It is important to note that, despite the perceived relevance of poisoning attacks, it is often remarkably difficult for an adversary to successfully run a poisoning campaign against classifiers operating on constraint-heavy tabular data, such as cybersecurity data --- like network flows or malware samples~\cite{severiExplanationGuidedBackdoorPoisoning2021}.
This is a well known issue in adversarial ML, illustrated in detail by~\cite{pierazziIntriguingPropertiesAdversarial2020} and often referred to as \emph{problem-space} mapping.
It stems from the complexity of crafting perturbations of the data points (in feature space) that induce the desired behavior in the victim model without damaging the structure of the underlying data object (problem space) necessary for it to be generated, parsed, or executed correctly.
When dealing with aggregated network flow data, these difficulties compound with the inherent complexity of handling multivariate tabular data consisting of heterogeneous fields.
To address these challenges, we design a novel methodology
based on ML explanation methods to determine important features for backdoor creation, and map them back into the problem space. 
Our methods handle complex dependencies in feature space, generalize to different models and feature representations, are effective at low poisoning rates (as low as 0.1\%), and generate stealthy poisoning attacks.

In summary, we make the following contributions:
\begin{enumerate*}[label=(\roman*)]
\item We develop a new strategy to craft clean-label, data-only, backdoor poisoning attacks against network traffic classifiers that are effective at low poisoning rates.
\item We show that our poisoning attacks work across different model types, classification tasks, and feature representations, and we comprehensively evaluate the techniques on several network traffic datasets used for malware detection and application classification.
\item We propose different strategies, including generative approaches based on Bayesian networks, to make the attacks inconspicuous and blend the poisoned data with the underlying training set.
\end{enumerate*}
To ensure reproducibility, we evaluate our techniques on publicly available datasets, and all the code used to run the experiments in the paper will be released upon publication.

\section{Background and Related Work}

\vspace{5pt}
\noindent\textbf{Machine Learning for Threat Detection.} Machine learning methods have been successfully used to detect several cyber security threats, including: malicious domains~\cite{antonakakis2010building,segugio,antonakakis11,MADE,CELEST}, command-and-control communication between attackers and compromised hosts~\cite{nelms2013execscent,MADE}, or malicious binaries used by adversaries for distributing exploit code and botnet
commands~\cite{nazca_ndss2014,tamersoy14}. 
Several endpoint protection products~\cite{microsoftdefender,ibmqradar,fireeye2020Jun,xdr_solutions} are now integrating ML tools to proactively detect the rapidly increasing number of threats.

\vspace{5pt}
\noindent\textbf{Adversarial Machine Learning.}
We can identify two major categories of integrity attacks against ML classifiers: (1) evasion attacks, which occur at test time and consist in applying an imperceptible perturbation to test samples in order to have them misclassified, and (2) poisoning attacks, which influence the training process (either through tampering with the training dataset or by modifying other components of the training procedure) to induce wrong predictions during inference. 
For details on other adversarial ML techniques, we direct the reader to the standardized taxonomy presented in~\cite{opreaAdversarialMachineLearning2023}.

In this study, we are focusing on backdoor poisoning attacks, a particularly insidious technique in which the attacker forces the learner to associate a specific pattern to a desired target objective --- usually the benign class in cybersecurity applications. While backdoor poisoning does not impact the model's performance on typical test data, it leads to misclassification of test samples that present the adversarial pattern. 
Backdoor poisoning attacks against modern ML models were introduced by Gu et al.~\cite{guBadNetsEvaluatingBackdooring2019} in BadNets, where a small patch of bright pixels (the trigger pattern) was added to a subset of images at training time together with an altered label, to induce the prediction of a target class. Subsequently, Turner et al.~\cite{turner2019labelconsistent} and Shafahi et al.~\cite{shafahiPoisonFrogsTargeted2018} devised clean-label backdoor attacks which require more poisoning data samples to be effective, but relax some strong assumptions of previous threat models, making them significantly more applicable in security scenarios.

In cybersecurity, the earliest poisoning attacks were designed against worm signature generation~\cite{Perdisci2006Worm,Newsome2006Paragraph} and spam detectors~\cite{Nelson2008}.  More recently, a few studies have looked at packet-level poisoning via padding~\cite{holodnakBackdoorPoisoningEncrypted2022,ningTrojanFlowNeuralBackdoor2022}, feature-space poisoning in intrusion detection~\cite{apruzzeseAddressingAdversarialAttacks2019,Li2018EPD}, and label flipping attacks for IoT~\cite{Papadopoulos2021}. Severi et al.~\cite{severiExplanationGuidedBackdoorPoisoning2021}
proposed to use model interpretation techniques to generate clean-label poisoning attacks against malware classifiers. Their strategies are applicable to security datasets whose records are independent such as individual files or Android applications, which present a direct mapping from feature space to problem space. In contrast, our study explores attacks trained on network traffic, where multiple sequential connections are translated into one single feature-space data point; in this setting, inverting triggers from feature to problem space becomes particularly difficult due to data dependencies. 

\vspace{5pt}
\noindent\textbf{Model Interpretation Techniques.}
With the proliferation and increase in complexity of ML models the field of explainable machine learning, focused on understanding and interpreting their predictions, has seen a substantial increase in popularity over recent years. 
We are particularly interested in model-agnostic interpretability techniques, which can be applied to any model. Linardatos et al.~\cite{Linardatos2021} provide a comprehensive taxonomy of these methods, and conclude that, among the black-box techniques presented, Shapley Additive explanations (SHAP)~\cite{SHAP,Lundberg2020} is the most complete, providing explanations for any model and data type both at a global and local scale.
SHAP is a game-theory inspired method, which attempts to quantify how important each feature is for a classifier's predictions. SHAP improves on other model interpretation techniques like LIME~\cite{ribeiro2016should}, DeepLIFT~\cite{shrikumar2017learning} and Layer-Wise Relevance Propagation~\cite{binder2016layer}, by introducing a unified measure of feature importance that is able to differentiate better among output classes.

In this study, we also experiment  with Gini index~\cite{Gastwirth72} and information gain~\cite{Koller96,LEE2006} -- two of the most popular splitting algorithms in decision trees. A decision tree is built recursively, by choosing at each step the feature that provides the best split. Thus, the tree offers a natural interpretability, and a straightforward way to compute the importance of each feature towards the model's predictions.

\vspace{5pt}
\noindent\textbf{Preserving Domain Constraints.}
Functionality-preserving attacks on network traffic have mostly looked at evasion during test time, rather than poisoning. For instance, Wu et al.~\cite{Wu2019Evading} proposed a packet-level evasion attack against botnet detection, using reinforcement learning to guide updates to adversarial samples in a way that maintains the original functionality.
Sheatsley et al.~\cite{Sheatsley2021DomainConstraints} study the challenges associated with the generation of valid adversarial examples that abide domain constraints and develop techniques to learn these constraints from data. Chernikova et al.~\cite{FENCE} design evasion attacks against neural networks in constrained environments, using an iterative optimization method based on gradient descent to ensure valid numerical domain values. With our constraint-aware problem-space mapping, which also takes into account dependencies in network traffic,  we delve one step further into the challenging issue of designing functionality-preserving attacks.  

Significant advances have been made recently with respect to generating multivariate data. 
Modern tabular data synthesizers of mixed data types leverage the power of generative adversarial networks~\cite{CTGAN_Xu2019,engelmann2021,Fan2020,CTAB-GAN-zhao21a,faketables2019} and diffusion models~\cite{kotelnikovTabDDPMModellingTabular2022} to create realistic content from the same distribution as the original data. Among the different frameworks, FakeTables~\cite{faketables2019} is the only attempt at preserving functional dependencies in relational tables. However, its evaluation is limited to Census and Air Carrier Statistics datasets, and its ability to capture more complex relationships between variables is unclear.   

In this work, we model conditional dependencies in the traffic using Bayesian networks -- a common choice for generating synthetic relational tables~\cite{deleu2022bayesian,Rezende2014,Heckerman2008,Kaur2020Bayesian,Young2009Bayesian}. Bayesian networks offer increased transparency and computational efficiency over more complex generative models like generative adversarial networks~\cite{Kaur2020Bayesian}. We believe this is an important advantage is our setting, which deals with large volumes of network traffic featuring multiple variables (e.g., log fields).
In cybersecurity, Bayesian networks have also been used to learn traffic patterns and flag potentially malicious attempts in intrusion detection systems~\cite{Muñoz2018Bayesian,Xu2010BayesianIDS,Devarakonda2012Bayesian,Jabbar2017BayesianIDS}.

\section{Threat Model}
\label{sec:threat}

\vspace{5pt}
\noindent\textbf{Adversary's Capabilities.}
Recent work analysing the training time robustness of malware classifiers~\cite{severiExplanationGuidedBackdoorPoisoning2021,yangJigsawPuzzleSelective2023} pointed out that the use of ever larger quantities of data to train effective security classifiers inherently opens up the doors to data-only poisoning attacks, especially in their more stealthy clean-label~\cite{turnerCleanLabelBackdoorAttacks2018,shafahiPoisonFrogsTargeted2018} variants where the adversary does not control the label of the poisoned samples.
Thus, in this work, we constrain the adversary to clean-label data-only attacks. 
This type of setup moves beyond the classic threat model proposed by Gu et al.~\cite{guBadNetsEvaluatingBackdooring2019} and adopted by other research~\cite{ningTrojanFlowNeuralBackdoor2022,Liu2018Trojaning,Chen2017Backdoor}, where the adversary was able to tamper with not only the content of the training points but also the corresponding ground-truth labels.
Here, instead, by disseminating innocuous looking ---but adversarially crafted--- data, the adversary is able to indirectly tamper with a small, yet effective, percentage of the training set and induce the desired behavior in the learned model.
To design the trigger, the adversary requires access to a small amount of clean labeled data, $D_a$, from a similar distribution as the victim's training data. In our experiments, we partition the test set in two disjoint sets of 85\% and 15\% of the points respectively, and supply the adversary with the smaller one.

We consider an adversary who has query-only access to the machine learning classifier. This allows the attacker to use the SHAP explanation technique to compute feature importance coefficients, but it prevents any form of inspection of model weights or hidden states. 
This scenario is very common for deployed models, as they often undergo periodical re-training but are only accessible behind controlled APIs.
Interacting with a victim system, however, always imposes a cost on the attacker, whether in terms of actual monetary expenses for API quotas, or by increasing the risk of being discovered.
Motivated by this observation, we also explore the use of model interpretation methods that do not require any access to the classifier, but instead leverage proxy models on local data (i.e., information gain and Gini coefficients), and can be used even when the model is not subject to re-training cycles.
Several previous studies on training time attacks~\cite{ningTrojanFlowNeuralBackdoor2022,Liu2018Trojaning} relax the model access constraints, assuming an adversary can train a ML classifier and provide it to the victim through third-party platforms such as Machine Learning as a Service (MLaaS)~\cite{Ribeiro2015MLaaS}. 
However, we believe that this threat model is rapidly becoming obsolete, at least in the cybersecurity domain, due to the push for stricter cyber hygiene practices from security vendors, including the reluctance to trust third-party model providers and MLaaS platforms~\cite{threat_modeling2022,Philipp2021MLaas}.

\vspace{5pt}
\noindent\textbf{Adversary's Objective.}
The main objective of the adversary is to acquire the ability to consistently trigger desired behavior, or output, from the victim model, after the latter has been trained on the poisoned data.
In this study, we focus on the binary class scenario (0/1), where the goal is reified into having points of a chosen \emph{victim} class being mis-labeled as belonging to the \emph{target} class, when carrying a backdoor pattern that does not violate the constraints of the data domain. For instance, in the benign/malicious case, the adversary attempts to have malicious data points mis-classified as benign, where ``benign'' represents the target class.

\vspace{5pt}
\noindent\textbf{Adversary's Target.}
We select two representative ML classifier models as target for our attacks: Gradient Boosting decision trees, and Feed-forward Neural Networks. Both of these models have been widely-used in intrusion detection for classifying malicious network traffic, with decision trees often preferred in security contexts due to their easier interpretation~\cite{Jacobs2022Emperor}.
We study two use cases of network traffic classifiers: (1) detection of malicious activities, and (2) application classification. 

\begin{table}[ht]
\centering
\caption{Network data format. Our data is represented by connection logs (``conn.log'' files) extracted with the Zeek monitoring tool from publicly-available packet-level PCAP files.}
\small
{\def\arraystretch{1.1}
\begin{tabular}{ll}
\hline
\textbf{Name} & \textbf{Description} \\ \hline
orig\_ip, resp\_ip             & Source and destination IP address  \\
orig\_p, resp\_p           & Source and destination port  \\ 
proto                   &Transport Protocol (e.g., TCP, UDP, or ICMP) \\
service         & Application protocol (e.g., ssh, dns, etc.)\\
ts          & Timestamp -- the connection start time\\
duration        & Duration of connection \\
orig\_pkts, resp\_pkts      & Number of transmitted packets \\
orig\_bytes, resp\_bytes    & Number of payload bytes \\
conn\_state         & Connection state, assessing whether the   \\
         & connection was established and terminated \\
        & normally (13 different states) \\
\hline 
\end{tabular}
}
\label{tab:conn_data}
\end{table}

\vspace{5pt}
\noindent\textbf{Data Format.}
In our threat model, network traffic consists of connection logs (``conn.log'' files), which are extracted from packet-level PCAP files using the Zeek monitoring tool. The Zeek log fields used in our study are described in \Cref{tab:conn_data}, and include port, IP address, protocol, service, timestamp, duration, packets, payload bytes and connection state.
Thus, the input data is tabular and multivariate, consisting of multiple log fields in either numeric format (e.g., bytes, packets, etc.) or categorical format (e.g., connection state, protocol, etc.). 
A data point in this domain is represented by a \emph{sequence} of raw log records grouped together. This \emph{problem-space} data point is mapped into a corresponding \emph{feature-space} data point through various aggregation techniques applied over the log field values.

\begin{table}[ht]
\centering
\caption{Statistical features aggregated over connection logs within each data point grouping.  The grouping is comprised of connections within 30-sec time windows, aggregated separately for each \emph{internal} IP and destination port within the time window. Note that the internal IP versus external IP distinction pertains to the subnet, not to the two ends of the connection (source/destination).}
\small
{\def\arraystretch{1.1}
\begin{tabular}{ll}
\hline
\textbf{Field} & \textbf{Description} \\ \hline
\multicolumn{2}{c}{Aggregation Key:} \\
\multicolumn{2}{c}{(time window, internal IP, destination port)} \\ \hline
proto & Count of connections per transport protocol\\
conn\_state     & Count of connections for each conn\_state \\
orig\_pkts, resp\_pkts  & Sum, min, max over packets\\
orig\_bytes, resp\_bytes  & Sum, min, max over bytes \\
duration   & Sum, min, max over duration \\\hline
\multicolumn{2}{c}{Aggregation Key: (time window, internal IP)} \\ \hline
ip     &Count of distinct \emph{external} IPs \\
resp\_p   & Count of distinct destination ports \\
\hline 
\end{tabular}
}
\label{tab:features}
\end{table}

\vspace{5pt}
\noindent\textbf{Feature Representation.}
We study two standard and widely adopted feature mapping techniques: (1) aggregation, to produce statistical features, and (2) embeddings--- using auto-encoders to automatically generate feature vectors. 
Traffic statistics have multiple applications in network monitoring and security~\cite{mooreDiscriminatorsUseFlowbased2005,yangFeatureExtractionNovelty2021}, which require dealing with large volumes of data. For instance, distinct count metrics are used to identify scanning attacks, while volume metrics or traffic distributions over port numbers and IP address ranges are utilized in anomaly detection~\cite{Burkhart2010FeatureStatistics}.  We use similar aggregation methods with previous works~\cite{Burkhart2010FeatureStatistics, ongunPORTFILERPortLevelNetwork2021}, to derive statistics of connections. The statistical features used in our study are described in \Cref{tab:features}, and include traffic volume by internal IP (in bytes and packets) within a 30-sec time window, connection counts by transport protocol, connection counts by state, etc.

Recent literature also features a variety of approaches for network traffic classification based on auto-encoders~\cite{mirskyKitsuneEnsembleAutoencoders2018a,yangFeatureExtractionNovelty2021, heDeepFeatureBasedAutoencoderNetwork2021, dangeloNetworkTrafficClassification2021}.
Auto-encoders are unsupervised models that learn to reconstruct the training data.
They are often used either for anomaly detection or to learn high level features to use in downstream classifiers.

\section{Attack Strategy}

The formulation of an appropriate trigger pattern is a fundamental aspect of backdoor poisoning attacks 
The inherent intricacies of network traffic ---feature dependencies, multiple data modalities--- makes it particularly challenging to ensure that the trigger is mapped correctly to realizable actions in problem space~\cite{pierazziIntriguingPropertiesAdversarial2020}.
This is a stark difference with the image domain, where the backdoor trigger can be extremely simplistic, such as a bright colored square~\cite{guBadNetsEvaluatingBackdooring2019}.

There are three key requirements that characterize a feasible poisoning attack: 
\begin{enumerate*}[label=(\roman*)]
    \item To be effective, the trigger should be easy to associate to the target class by the victim model.
    \item The injected pattern should appear inconspicuous, so as to avoid detection by potential human or automated observers.
    \item The perturbations induced by the injection of the trigger pattern should not affect data validity. 
\end{enumerate*}
While the first two requirements are generic to any backdoor attack, the third one translates to additional constraints on adversarial actions in the network domain, specifically:
\begin{enumerate*}[label=(\roman*), noitemsep]
\item The adversary can only insert traffic, but not modify or remove existing traffic.
\item Data semantics and dependencies need to be preserved, such as value restrictions on specific fields (e.g., upper/lower bounds on packet length), feature correlations (e.g., protocols use specific ports), etc.
\item The injected pattern needs to handle multiple data types, i.e., numeric and categorical.\end{enumerate*}

\subsection{Crafting the Poisoning Data}
\label{sec:strategy_poison}

\begin{figure}
    \centering
    \includegraphics[width=0.95\linewidth]{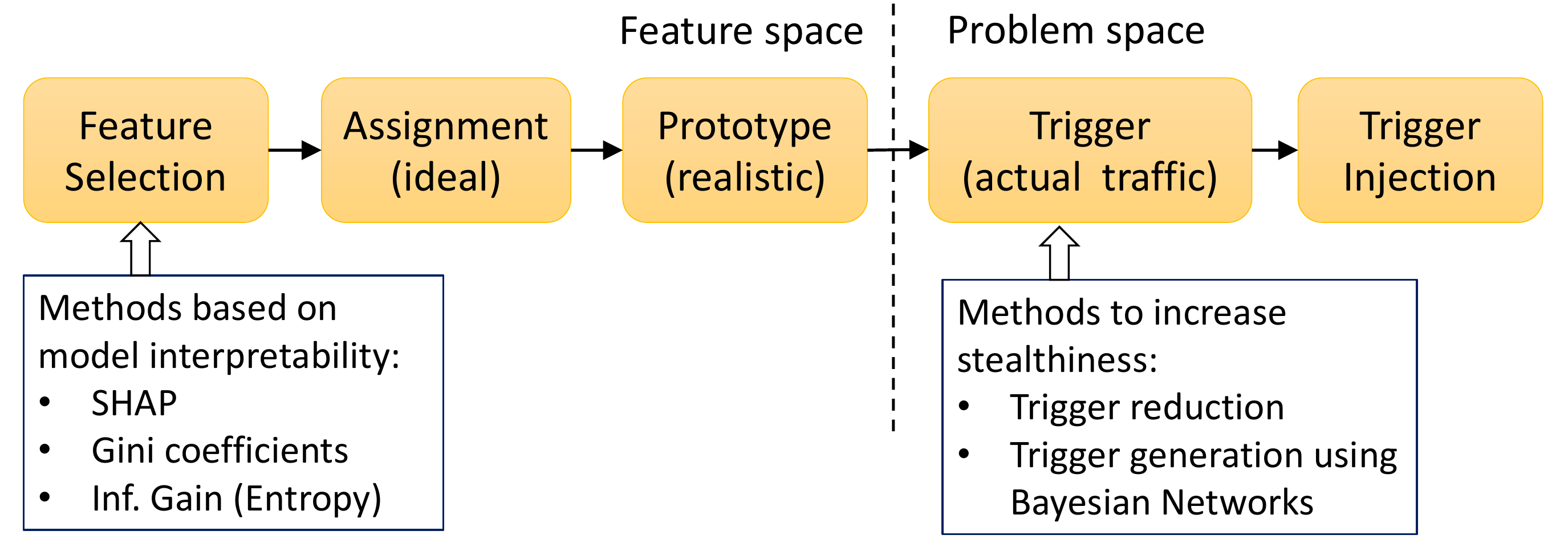}
    \caption{Pipeline for poisoning network flow classifiers.}
    \label{fig:attack_strategy}
\vspace{-5pt}
\end{figure}

To address the above challenges, we design a novel methodology that leverages insights from explanation-based methods to determine important features in feature space, then map them back to constraint-aware triggers in problem space. The mapping can be done via: (i) poisoning attacks using connections directly extracted from malicious traffic; (ii) poisoning attacks with reduced footprint; (iii) generative Bayesian models to increase attack stealthiness. Our attack strategy, illustrated in Figure~\ref{fig:attack_strategy}, consists of five main phases:
\begin{enumerate}[label=(\Roman*)]
     \item Select a subset of features that are most important for the class that the  adversary wishes to misclassify using model explanation techniques;
    \item Find an ideal trigger in feature space --- we call this an \emph{assignment};
    \item Find a data point that best approximate the ideal trigger values --- this will be our \emph{prototype} trigger;
    \item Identify a set of real connections that induce the values observed in the prototype --- this set of connections will be our actual \emph{trigger};
    \item Inject the trigger in points of the target class, potentially trying to minimize its conspicuousness.
\end{enumerate}

\paragraph{Phase I} We first identify the most relevant features for the class to be misclassified.
Our goal is to leverage highly informative features to coerce the model into associating the trigger pattern with the target class.
There are a variety of techniques from the field of model interpretability used to estimate the effect of specific features towards the classifier's decision.
We start by adapting the SHAP-based technique from~\cite{severiExplanationGuidedBackdoorPoisoning2021} to the network domain.
Here, SHAP values are computed for a subset of points to which the adversary has access to, and their contributions summed per-feature, to identify the ones most contributing to each class.
This approach has the advantage of being model agnostic, allowing us to estimate feature importance coefficients for any possible victim model.
Unfortunately, it also assumes the adversary is able to perform a possibly large number of queries against the victim model.
To address this potential limitation, we also evaluate the effect of selecting the important features through more indirect ways.
In particular we can leverage the \emph{information gain} and \emph{Gini coefficient} metrics used in training decision trees, to estimate the global contributions of each feature.
The attentive reader will notice here that the approaches we mentioned to estimate feature importance are quite different.
This is intentional, and it highlights the modularity of this component.
As long as the adversary is capable of obtaining global estimates of feature importance scores, they can use them to guide the attack.
Moreover, with potential future discoveries in the, extremely active, field of model interpretation, novel methods could be used to improve the effectiveness of this attack.

\paragraph{Phase II} 
Once the subset of important features is selected, we can proceed to find a suitable \emph{assignment} of values.
To be consistent with real traffic constraints, we need to ensure that the values that we select represent information that can be easily added to data points of the non-target class, by injecting new connections, without having to remove existing connections. 
Thus, we select values that correspond to the top $t^{th}$ percentile of the corresponding features for non-target class points; in practice, setting this parameter to $95^{th}$ percentile performed well in our experiments. 
Note that the non-target class points are generated by software under the control of the adversary, and therefore we assume they have access to a collection of log rows that represent those connections.

\paragraph{Phase III} Armed with the desired \emph{assignment} for the selected features, we can proceed to identify an existing data point that approximates these ideal trigger values.
To find it, in our first attack we leverage a \emph{mimicry} method to scan the non-target (e.g., malicious) class samples and isolate the one with the lowest Euclidean distance from the assignment, in the subspace of the selected features.
We call this point in feature space the trigger \emph{prototype}.

\paragraph{Phase IV} Up until this point, the process was working completely in feature space. 
Our explicit goal, however, is to run the attack in problem space.
So the next step in the attack chain is to identify, in the attacker's dataset, a contiguous subset of log connections that best approximate the \emph{prototype}. Enforcing that the selected subset is contiguous ensures that temporal dependencies across log records are preserved.
This subset of connections represents the actual \emph{trigger} that we will use to poison the target-class training data.

\paragraph{Phase V} Finally, it is time to inject the trigger in the training data.
This step is quite straightforward, as it only requires the software under control of the adversary, to execute the \emph{trigger} connections in the specified order. We next describe two strategies for increasing trigger stealthiness before injection.

\subsection{Increasing Attack Stealthiness}
\label{sec:method_stealthy}
Beyond the basic objective of maximizing attack success, the adversary may have the additional goal of minimizing the chance of being detected.
To achieve this secondary goal, the adversary may wish to slightly alter the trigger before injecting it in the training data.
In particular, we study two strategies: (1) trigger size reduction and (2) trigger generation using Bayesian models.

\vspace{5pt}
\noindent\textbf{Trigger size reduction.}
The first strategy consists of minimizing the trigger footprint, by removing all the connections that are not strictly necessary to achieve the values specified in the \emph{prototype} for the subset of important features (such as connections on other ports). 
We then select the smallest subset of contiguous connections that would produce the desired values for the selected features.

\vspace{5pt}
\noindent\textbf{Trigger generation using Bayesian networks.}
The second strategy aims at reducing the conspicuousness of the trigger by blending it with the set of connections underlying the data point where it is embedded.
To this end, we generate the values of the log fields corresponding to \emph{non-selected} features in the backdoor to make them appear closer to values common in the target-class natural data $\in D_a$. Note that fields influencing the selected (important) features will \emph{not} be modified, because they carry the backdoor pattern associated with the target class.
Our generative approach leverages Bayesian networks, a widely-used probabilistic graphical model for encoding conditional dependencies among a set of variables, and deriving realistic samples of data~\cite{deleu2022bayesian,Rezende2014,Heckerman2008}. 
Bayesian networks consist of two parts: (1) structure -- a directed acyclic graph (DAG) that expresses dependencies among the random variables associated with the nodes, and (2) parameters -- represented by conditional probability distributions associated with each node. 

\vspace{5pt}
\emph{Structure.} Given our objective to synthesize realistic log connections (in problem space) that lead to the feature-space prototype, we construct a directed acyclic graph $G=(V, E)$ where the nodes $x_i \in V$ correspond to fields of interest in the connection log and the edges $e_{ij} \in E$ model the inter-dependencies between them. 
We explore field-level correlations in connection logs using two statistical methods that have been previously used to study the degree of association between variables~\cite{Kitson2023Bayesian}: the correlation matrix and the pairwise normalized mutual information. In our experiments, both methods discover similar relationships in $D_a$, with the mutual information approach bringing out additional inter-dependencies. 
Note that we are not interested in the actual coefficients, rather, in the associational relationships between variables. Thus, we extract the strongest pairwise associations, and use them in addition to domain expertise to guide the design of the DAG structure. 
For instance, there is a strong relationship between the number of response packets and source packets (resp\_pkts $\leftrightarrow$ orig\_pkts); between the protocol and the response port (proto $\leftrightarrow$  resp\_p); between the connection state and protocol (conn\_state $\leftrightarrow$ proto), etc.

There is a large body of literature on learning the DAG structure directly from data. We point the interested reader to a recent survey by Kitson et al.~\cite{Kitson2023Bayesian}. However, computing the graphical structure remains a major challenge, as this is an NP-hard problem, where the solution space grows super-exponentially with the number of variables. Resorting to a hybrid approach~\cite{Kitson2023Bayesian} that incorporates expert knowledge is a common practice that alleviates this issue.
The survey also highlights the additional complexity in modeling the DAG when continuous variables are parents of discrete ones, and when there are more than two dependency levels in the graph.

\begin{figure} [t]
    \centering
    \includegraphics[width=0.9\linewidth]{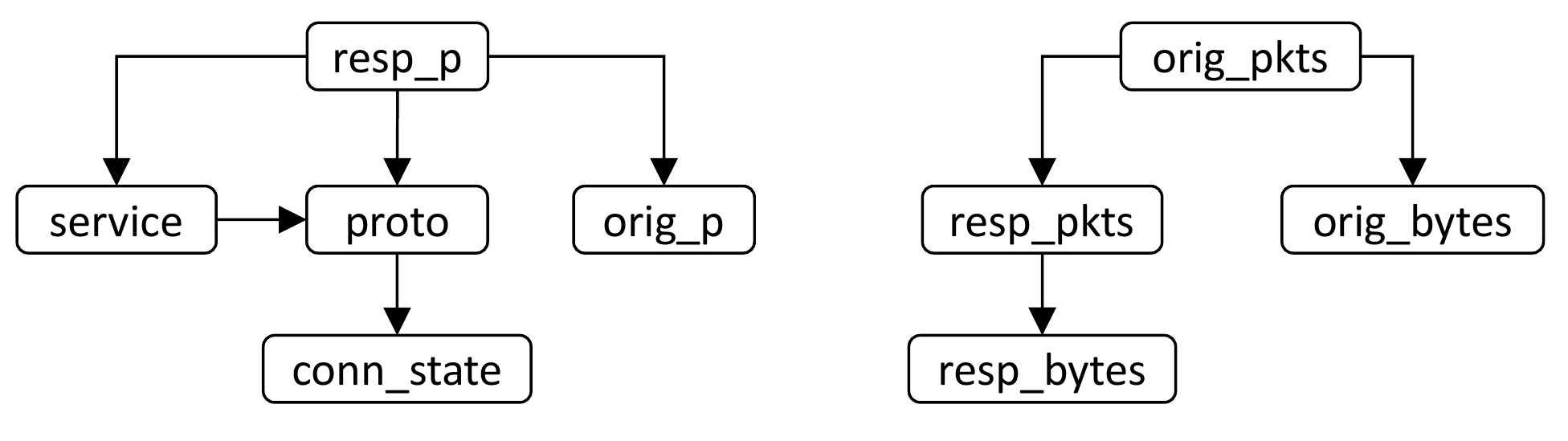}
    \caption{Directed Acyclic Graph (DAG) representing the inter-dependencies between log connection fields.}
    \label{fig:DAG}
\vspace{-5pt}
\end{figure}

\begin{figure} [t]
    
    \begin{subfigure}[b]{0.48\textwidth}
    \centering
    \includegraphics[width=0.9\linewidth]{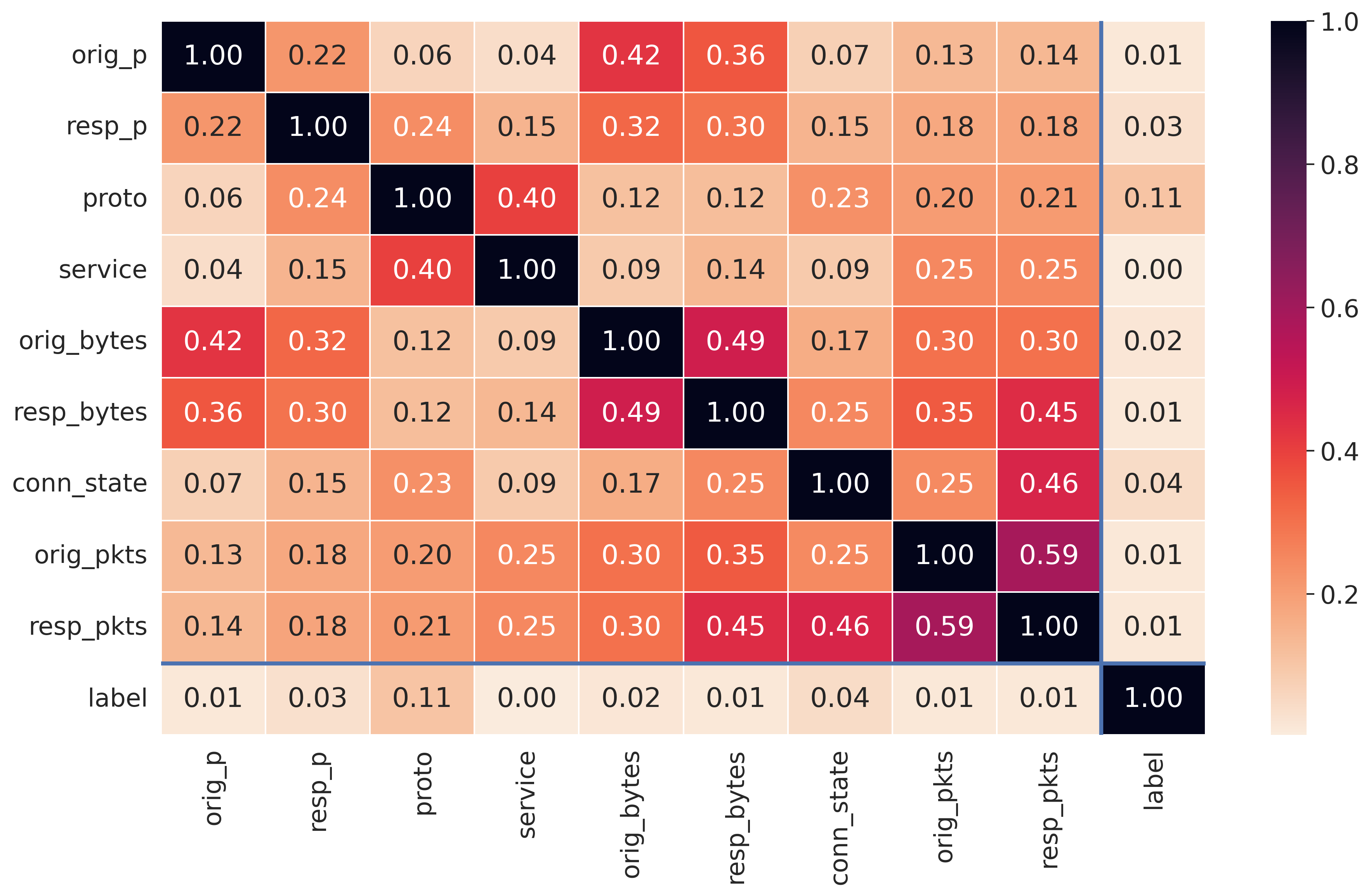}
    \vspace{-5pt}
    \caption{Mutual information on \emph{clean data}, computed on the adversary's dataset.}
    \label{fig:neris_adv_mi}
    \end{subfigure}
    
    \vspace{10pt}
    
    \begin{subfigure}[b]{0.48\textwidth}
    \centering
    \includegraphics[width=0.9\linewidth]
    {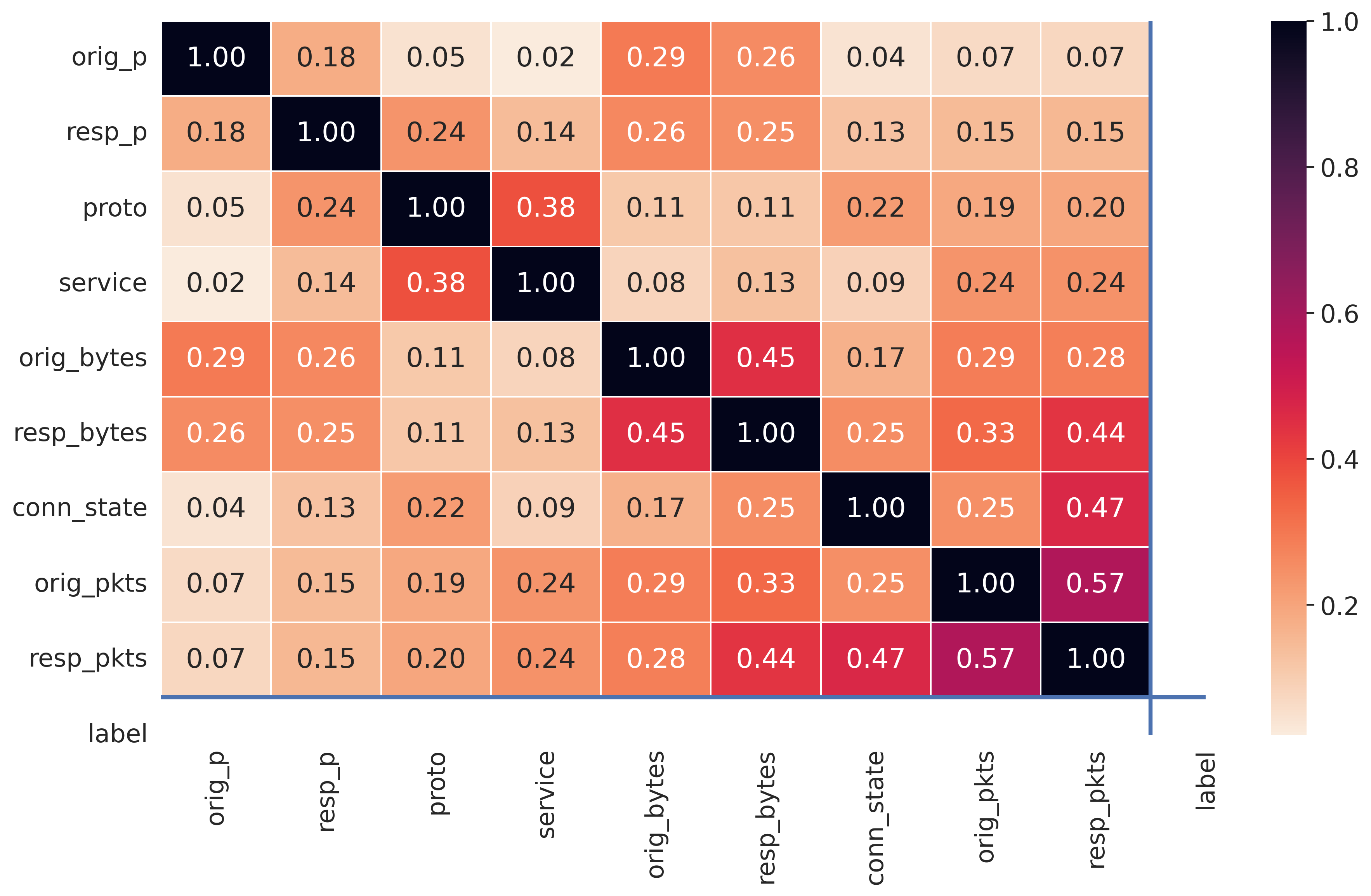}

    \vspace{-5pt}
    \caption{Mutual information on the \emph{poisoned training dataset} }
        \label{fig:neris_pois_mi}
    \end{subfigure}  
    
    \label{fig:neris_mi}
    \caption{Mutual information comparison on clean and poisoned data. Showing associations between relevant fields of the \textit{conn.log} file for CTU-13.} 
    
\end{figure}

Based on the above considerations, we design the direct acyclic graph presented in \Cref{fig:DAG}. 
For practical reasons, we filter out some associations that incur a high complexity when modeling the conditional probability distributions. To ensure that the generated traffic still reflects the inter-dependency patterns seen in the data, we inspect the poisoned training dataset using the same statistical techniques (correlation matrix and mutual information). We include the mutual information matrix on the clean adversarial dataset (\Cref{fig:neris_adv_mi}) and on the training dataset poisoned with the Generated trigger method (\Cref{fig:neris_pois_mi}), to show that the associational relationships between variables are preserved after poisoning (though the actual coefficients may vary).

\vspace{5pt}
\emph{Parameters.} Bayesian networks follow the local Markov property, where the probability distribution of each node, modeled as a random variable $x_i$, depends only on the probability distributions of its parents. Thus, the joint probability distribution of a Bayesian network consisting of $n$ nodes is represented as:
$p(x_1, x_2,  \cdots, x_n) = \prod_{i=1}^n p(x_i | x_{P_i})$,
where $P_i$ is the set of parents for node $i$, and the conditional probability of node $i$ is expressed as $p(x_i | x_{P_i})$. 

\begin{figure} [ht]
\centering
\includegraphics[width=0.95\linewidth]{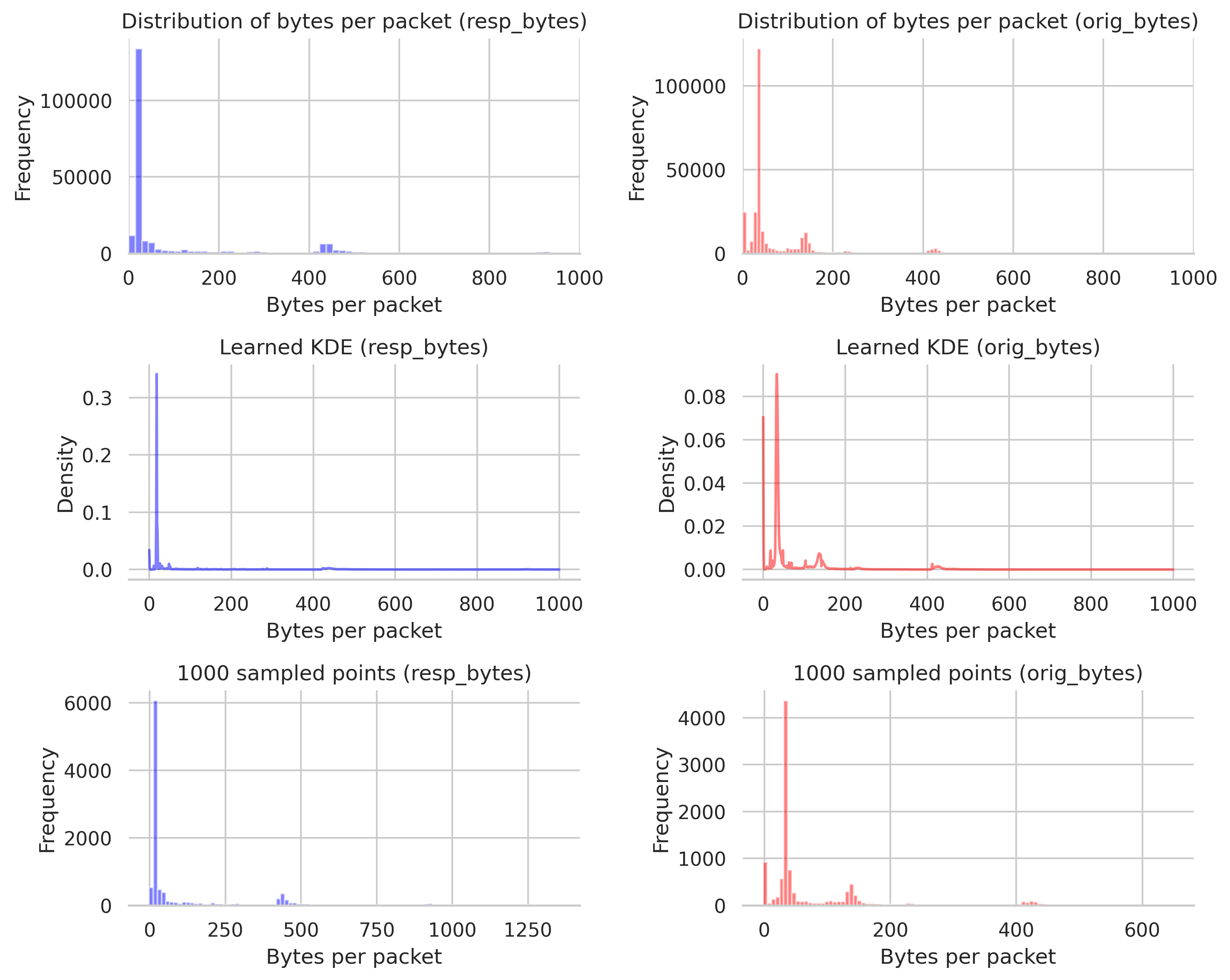}
\caption{Modeling the bytes distribution for responder (left side) and originator (right side): From top to bottom, the figures show: distribution of byte counts per packet, learned KDEs, and sampled data from the learned distributions.}
\label{fig:neris_kde_bpp}
\end{figure}

\vspace{5pt}
\emph{Sampling.}
The DAG is traversed in a hierarchical manner, one step at a time, as a sequential decision problem based on probabilities derived from the data, with the goal of generating a realistic set of field-value assignments. The value assignments for nodes at the top of the hierarchy are sampled independently, from the corresponding probability distribution, while the nodes on lower levels are conditioned on parent values during sampling. 
We compute the conditional probabilities of categorical fields (e.g., ports, service, protocol, connection state), and model numerical fields (e.g., originator/responder packets and bytes) through Gaussian kernel density estimation (KDE).
An example of the KDE learned from the data, and used to estimate the number of exchanged bytes between a source (originator) and a destination (responder), given the number of packets, is presented in \Cref{fig:neris_kde_bpp}.

\begin{table*}[t]
\centering
\caption{Sampling method for each dependency described in the DAG from \Cref{fig:DAG}. In this example, we assume that the most important features correspond to protocol and port; their values (TCP protocol on port 80) have been determined in Phase II of our strategy. Here, our generative method samples the rest of the log field values. $D_a$ represents the attacker's dataset.}
\small
{\def\arraystretch{1.1}
\begin{tabular}{|>{\arraybackslash}p{1.2in}
|>{\arraybackslash}p{5in} |}

\hline
\textbf{Dependency} & \textbf{Sampling method} \\ \hline
1. resp\_p $\rightarrow$ service  & Select subset from attacker's data, $D_a$, with resp\_p = 80. Sample a value for service (S) according to the observed probabilities.\\
2. service $\rightarrow$ conn\_state  & Subset $D_a$ with proto = TCP and service = S. Sample conn\_state according to the observed probabilities.\\
3.  resp\_p $\rightarrow$ orig\_p & Subset $D_a$ with resp\_p = 80. Sample orig\_p according to the observed probabilities.\\
4. orig\_pkts & Sample a value for orig\_pkts from the KDE learned on $D_a$. \\
5.  orig\_pkts $\rightarrow$ resp\_pkts & Subset $D_a$ based on orig\_pkts. Learn the KDE for resp\_pkts from the subset. Sample resp\_pkts from the KDE. \\ 
6. orig\_pkts $\rightarrow$ orig\_bytes & Learn the KDE distribution $D_O$ of originator bytes-per-packet from $D_a$. Given previously sampled value for number of packets, orig\_pkts = $m$, sample and sum up $1, \cdots, m$ values from the distribution $D_O$. \\
7. resp\_pkts $\rightarrow$ resp\_bytes & Learn the KDE distribution $D_R$ of responder bytes-per-packet from $D_a$. Given previously sampled value for number of packets, resp\_pkts = $n$, sample and sum up $1, \cdots, n$ values from the distribution $D_R$. 
 \\
\hline 
\end{tabular}
}
\label{tab:sampling}
\vspace{-5pt}
\end{table*}

Given the complexity of sampling from hybrid Bayesian networks, we approximate the conditional sampling process with a heuristic, described in  \Cref{tab:sampling}. We consider an example where the log fields corresponding to the most important features have been set to the TCP protocol and responder port 80. Our generative method synthesizes values for the rest of the fields, in an attempt to make the trigger blend in with the target class.  
We show in our evaluation that the synthesized poisoning traffic is a good approximation of clean network traffic, both in terms of Jensen-Shannon distance between distributions (\Cref{sec:results_stealthy}) and preservation of field-level dependencies.

\section{Experimental Results}

\subsection{Experimental Setup}
In this section, we describe the datasets and performance metrics used in our evaluation. We also present the baseline performance of the target classifiers (without poisoning). 

\vspace{5pt}
\noindent\textbf{Datasets.} We used three public datasets commonly used in cybersecurity research for intrusion detection and application classification.

\paragraph{CTU-13 Neris Botnet:} We started our experimentation with the Neris botnet scenario of the well-known CTU-13 dataset~\cite{garciaEmpiricalComparisonBotnet2014}.
This dataset offers a window into the world of botnet traffic, captured within a university network and featuring a blend of both malicious and benign traffic.
Despite the sizeable number of connections ($ \approx 9*10^6$), the classes are extremely imbalanced, with a significantly larger number of benign than malicious data points. Note that the class imbalance is a common characteristic of security applications. 
The Neris botnet scenario unfolds over three capture periods. We use two of these periods for training our models, and we partition the last one in two subsets, keeping 85\% of the connections for the test set, and 15\% for the adversarial set, $D_a$.

\paragraph{CIC IDS 2018 Botnet:} From CTU-13, we moved to a recent dataset for intrusion detection systcheems, the Canadian Institute for Cybersecurity (CIC) IDS 2018
dataset~\cite{sharafaldinGeneratingNewIntrusion2018}.
We experimented with the botnet scenario, in which the adversary uses the Zeus and Ares malware packages to infect victim machines and perform exfiltration actions. This dataset includes a mixture of malicious and benign samples and is also heavily imbalanced.

\paragraph{CIC ISCX 2016 dataset:} 
This dataset contains several application traffic categories, such as chat, video, file transfer.
We leverage the CIC ISCX 2016 dataset~\cite{draper-gilCharacterizationEncryptedVPN2016} to explore another scenario where an adversary may affect the outcome via poisoning: detection of banned applications. For instance, to comply with company policies, an organization monitors its internal network to identify usage of prohibited applications. An adversary may attempt to disguise traffic originating from a banned application as another type of traffic.  We study two examples of classification tasks on the non-vpn traffic of this dataset: (1) \emph{File vs Video}, where we induce the learner to mistake video traffic flows as file transfer, and (2) \emph{Chat vs Video}, where the classifier mis-labels video traffic as chat communication.

\vspace{5pt}
\noindent\textbf{Performance Metrics.}
Similar to previous work in this area~\cite{severiExplanationGuidedBackdoorPoisoning2021,ningTrojanFlowNeuralBackdoor2022}, we are interested in the following indicators of performance for the backdoored model:
\begin{itemize} [topsep=2pt, itemsep=2pt]
\item \textit{Attack Success Rate (ASR)}. This is the fraction of test data points which are mis-classified as belonging to the target class. We evaluate this metric on a subset of points that have been \emph{previously correctly classified} by a clean model trained with the same original training data and random seed. 

\item \textit{Performance degradation on clean data}. This metric captures the side effects of poisoning, by evaluating the ability of the backdoored model to maintain its predictive performance on clean samples. Let $F_1^p$ be the F1 score of the poisoned model on the clean test set, and $F_1^c$ the test score of a non-poisoned model trained equally, the performance degradation on clean data at runtime is: $\Delta F_1 = |F_1^p - F_1^c|$.
\end{itemize}
Unless otherwise noted, all the results shown in the following sections are averages of five experiments with different random  seeds, reported with their relative standard deviations.

\vspace{5pt}
\noindent\textbf{Parameters.}
We define $p\%$ as the percentage of feature-space points of the training dataset that have been compromised by an adversary. Since the amount of poisoned points is generally a critical parameter of any poisoning attack, we measure the attack performance across multiple poison percentage values $p\%$ .  
At runtime, we randomly select a subset of test points to inject the trigger. Specifically, we select 200  points for the CTU-13 and CIC IDS 2018 datasets, and 80 for the CIC ISCX 2016 dataset (due its smaller size).

\vspace{5pt}
\noindent\textbf{Baseline Model Performance.}
As mentioned in our threat model, we consider two representative classifiers: a Gradient Boosting Decision Tree (GB), and a Feed Forward Neural Network (FFNN).
Note that we are not interested in finding the most effective possible learner for the classification task at hand, instead our focus is on selecting generic and widely adopted classifiers to showcase the adaptability of our attack strategy.
Baseline values for accuracy, F1 score, precision and recall of the classifiers are reported in \Cref{tab:base_perf}.

\begin{table}[t]
\centering
\caption{Base performance of the classifiers, avg. over 5 runs.}
\small
{\def\arraystretch{1.1}

\begin{tabular}{lllll} \hline
\textbf{Model} & \textbf{Accuracy} & \textbf{F1 score} & \textbf{Precision} & \textbf{Recall} \\ \hline
\multicolumn{5}{c}{CTU-13 Neris Botnet}                                                              \\
GB             & 0.999             & 0.959             & 0.996              & 0.925           \\
FFNN           & 0.999             & 0.927             & 0.971              & 0.887           \\ \hline
\multicolumn{5}{c}{CIC-IDS 2018 Botnet}                                                   \\
GB             & 0.999             & 0.994             & 0.993              & 0.995           \\
FFNN           & 0.999             & 0.995             & 0.999              & 0.991           \\ \hline
\multicolumn{5}{c}{ISCX 2016 File/Video}                                                           \\
GB             & 0.962             & 0.800             & 0.799              & 0.802           \\
FFNN           & 0.941             & 0.719             & 0.666              & 0.780           \\ \hline
\multicolumn{5}{c}{ISCX 2016 Chat/Video}                                                           \\
GB             & 0.936             & 0.901             & 0.928              & 0.875           \\
FFNN           & 0.947             & 0.919             & 0.939              & 0.900   \\  \hline       
\end{tabular}
}
\label{tab:base_perf}
\vspace{-5pt}
\end{table}

\subsection{Impact of Feature Selection}
  
\begin{figure}[] 
\begin{subfigure}[b]{0.45\textwidth}
    \centering
    \includegraphics[width=0.9\linewidth]{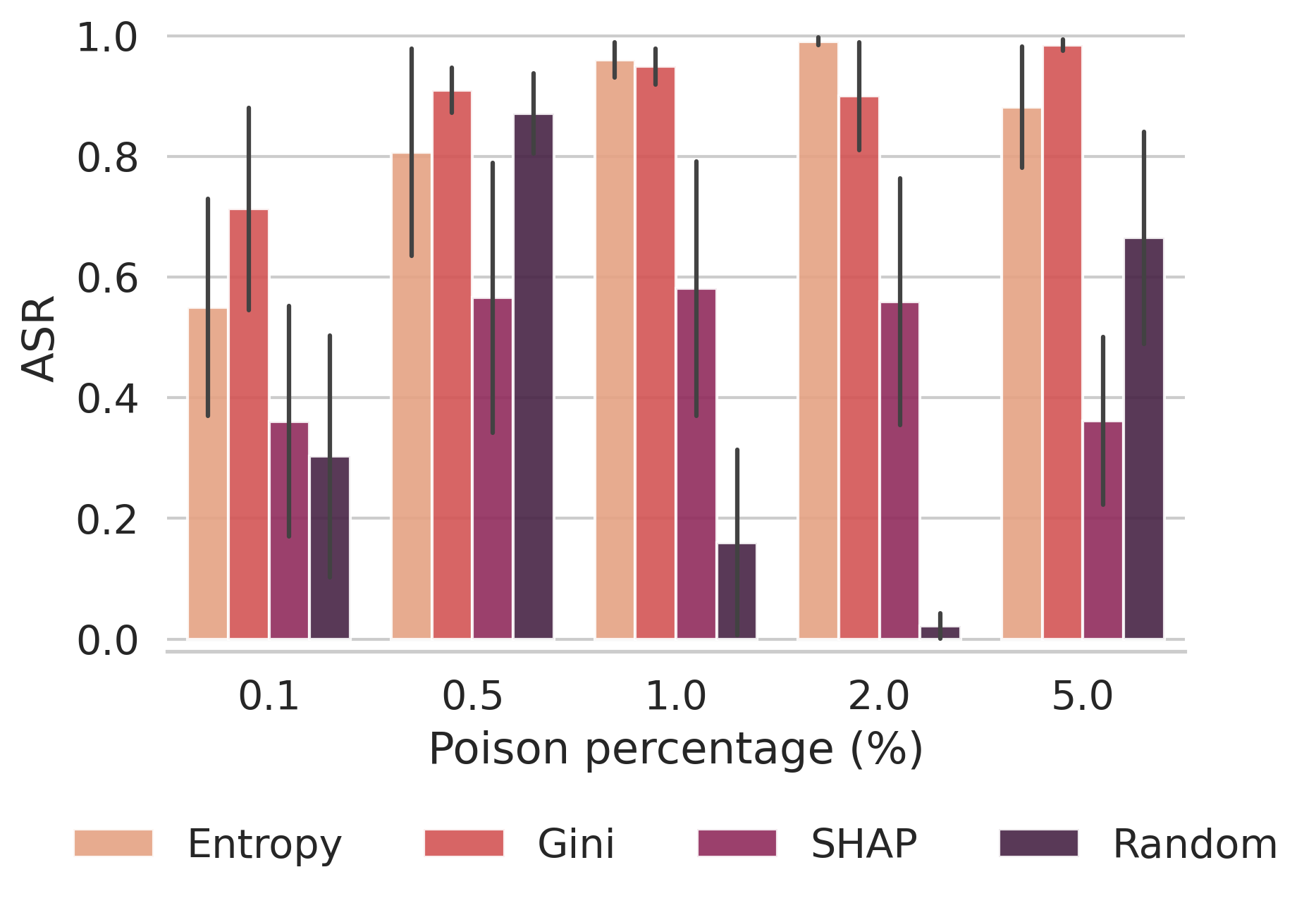}
    \caption{Gradient Boosting model}
    \label{fig:neris_gb_ft_asr}
\end{subfigure} 
\begin{subfigure}[b]{0.45\textwidth}
    \centering
    \includegraphics[width=0.9\linewidth]{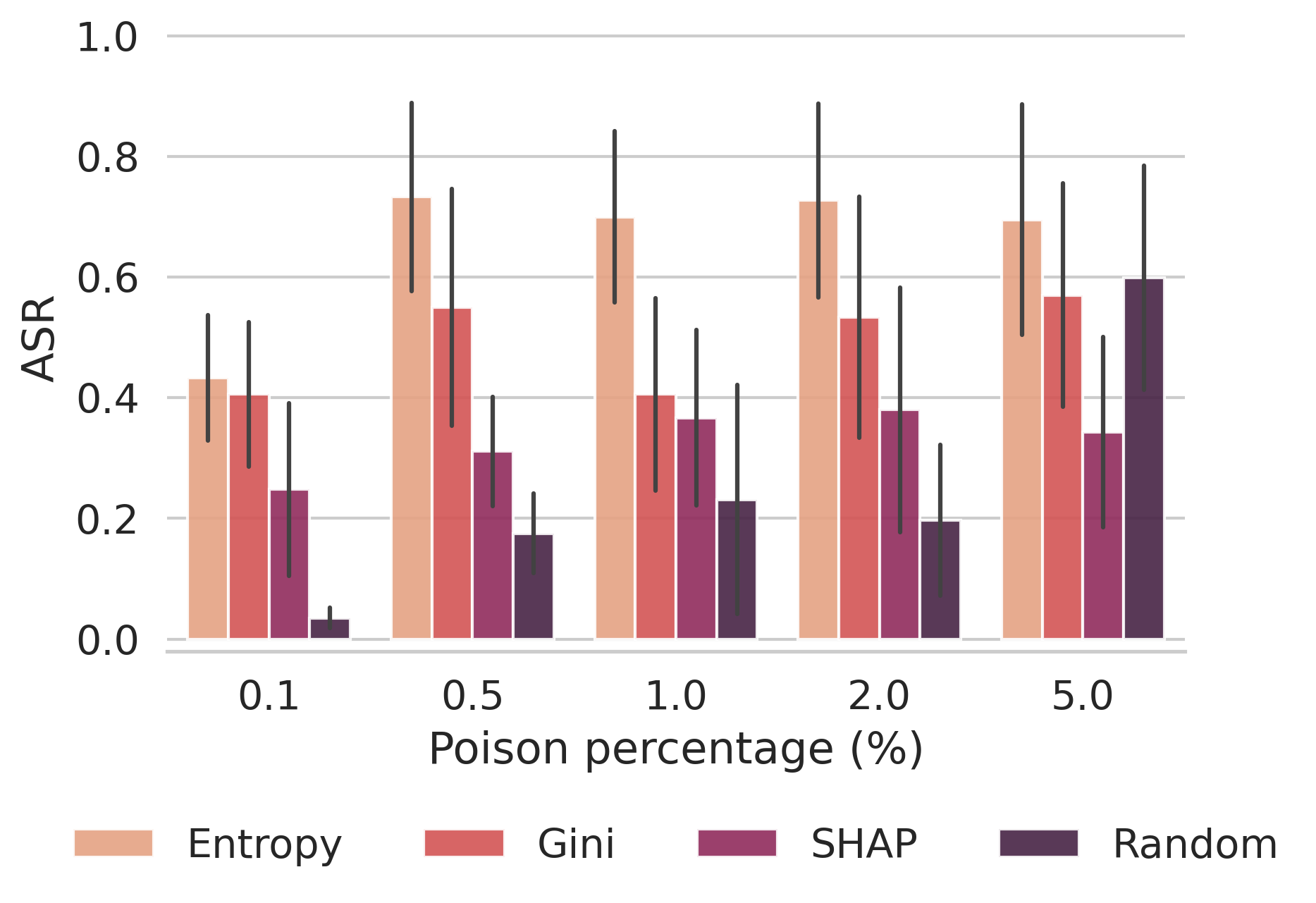}
    \caption{Feed-forward Neural Network model}
    \label{fig:neris_fn_ft_asr}
\end{subfigure}
    
\caption{Attack success rate (ASR) for the CTU-13 Neris Botnet scenario with different models and feature selection strategies.}
\label{fig:neris_asr}
\vspace{-5pt}
\end{figure}

Similar to the procedure reported in~\cite{severiExplanationGuidedBackdoorPoisoning2021}, our initial feature selection strategy revolved around computing local feature importance scores with SHAP and then aggregating them to obtain global indicators for each feature of the magnitude and direction of impact for each feature.
As mentioned in \Cref{sec:strategy_poison}, however, this approach has an important drawback: it requires to perform a potentially large number of queries against the victim classifier.
To obviate this issue, we also considered ways in which the adversary can extract feature importance estimates directly from their data subset, $D_a$. 
In practice, we experimented with fitting a Decision Tree on $D_a$, following either the Gini impurity (\emph{Gini}) or the information gain (\emph{Entropy}) criteria, and using the importance estimate given by the reduction of the criterion induced by the feature\footnote{Using the  implementation in Scikit-Learn \url{https://scikit-learn.org/stable/modules/generated/sklearn.tree.DecisionTreeClassifier.html}}.

The three feature selection strategies implemented (Entropy, Gini, SHAP) use the top eight most important features to design the trigger pattern, and are compared against Random, a baseline strategy that chooses the same number of features uniformly at random. Looking at the features selected by the different strategies, we generally observe that Entropy and Gini tend to assign scores that are strongly positive only for a very small number of features (typically 1-3), while SHAP scores are distributed more evenly. 
This observation, together with the desire to minimize the trigger footprint, informed our decision to select eight most relevant features.
We also experimented with different values of this parameter, halving and doubling the number of selected features, but we found that eight were sufficient to achieve satisfying success rates. 

\textit{Attack Success Rate:} We show the results of these experiments in \Cref{fig:neris_asr}. On average, we found the Entropy strategy to be the most successful against both classifiers on this dataset. The Random strategy leads to inconsistent results: occasionally, it stumbles upon useful features, but overall attacks relying on Random selection perform worse than attacks guided by the other feature selection methods.
\Cref{fig:neris_asr} also illustrates a major finding -- our attacks perform well even at very small poisoning rates such as 0.1\%, where they reach an attack success rate of up to 0.7 against the Gradient Boosting classifier. As expected, increasing the poisoning percentage leads to an increase in attack success rate; for instance, an ASR of 0.95 is obtained with Entropy at 1.0\% poisoning. 
This is interesting considering that previous works only considered larger poisoning rates (e.g, 2\% to 20\% in~\cite{liChronicPoisoningMachine2018a}, 20\% samples from nine (out of ten) non-target classes in~\cite{ningTrojanFlowNeuralBackdoor2022}).
We also notice that some of the variance in the ASR results can be attributed to a somewhat bimodal distribution.
This can be partially explained with differences in the resulting trigger sizes, with \Cref{fig:neris_entropy_trigger_size} highlighting the correlation between larger triggers and higher ASR.
We leave a more detailed analysis of the distribution of the ASR scores for future work.
The second interesting observation we can make, is that the SHAP strategy, while working well in some scenarios (especially for the application classification tasks in \Cref{sec:other_data}) does not, on average, lead to better results than estimating feature importance through proxy models (Entropy and Gini).
This makes the attack quite easier to run in practice, as it circumvents the necessity to run multiple, potentially expensive, queries to the victim model.

\textit{Performance degradation on clean data:} While these results show that the poisoned model is able to successfully misclassify \emph{poisoned} data, we also want make sure that the performance on \emph{clean} data is maintained. 
The average $\Delta F_1$ across poisoning rates and feature selection strategies in our experiments was below 0.037, demonstrating that the side effects of the attack are minimal.
The neural network model exhibits on average a slightly larger decrease when compared against the Gradient Boosting classifier, especially when the Entropy and Gini feature selection strategies are used.

\subsection{Attack Stealthiness}
\label{sec:results_stealthy}
Remaining undetected is an important factor in running a successful poisoning campaign.
Here, we study the impact of our two approaches for increasing attack stealthiness described in \Cref{sec:method_stealthy}: reducing the trigger size (\emph{Reduced} trigger) and generating the trigger connections using Bayesian networks (\emph{Generated} trigger). 
We start by analyzing the attack success with the different types of triggers, followed by a  quantitative comparison of their stealthiness in feature space (via anomaly detection), and in problem space (via the Jensen-Shannon distance).

\begin{figure*}[th]
    
    \begin{subfigure}[b]{\textwidth}
    \centering
    \includegraphics[width=0.7\linewidth]{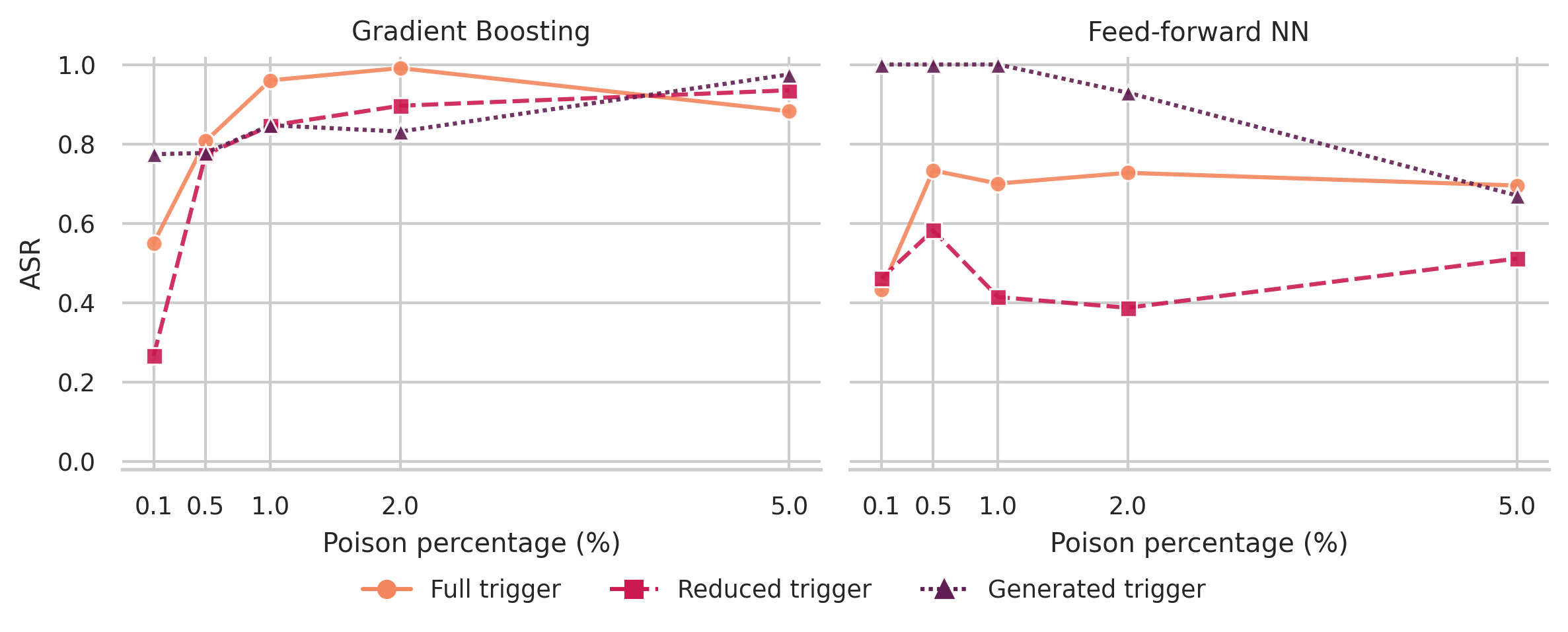}
    \caption{Comparison of attack success rates (ASR) as a function of poisoning percentage.}
    \label{fig:entropy_neris}
    \end{subfigure}

    \begin{subfigure}[b]{\textwidth}
    \centering
    \includegraphics[width=0.7\linewidth]{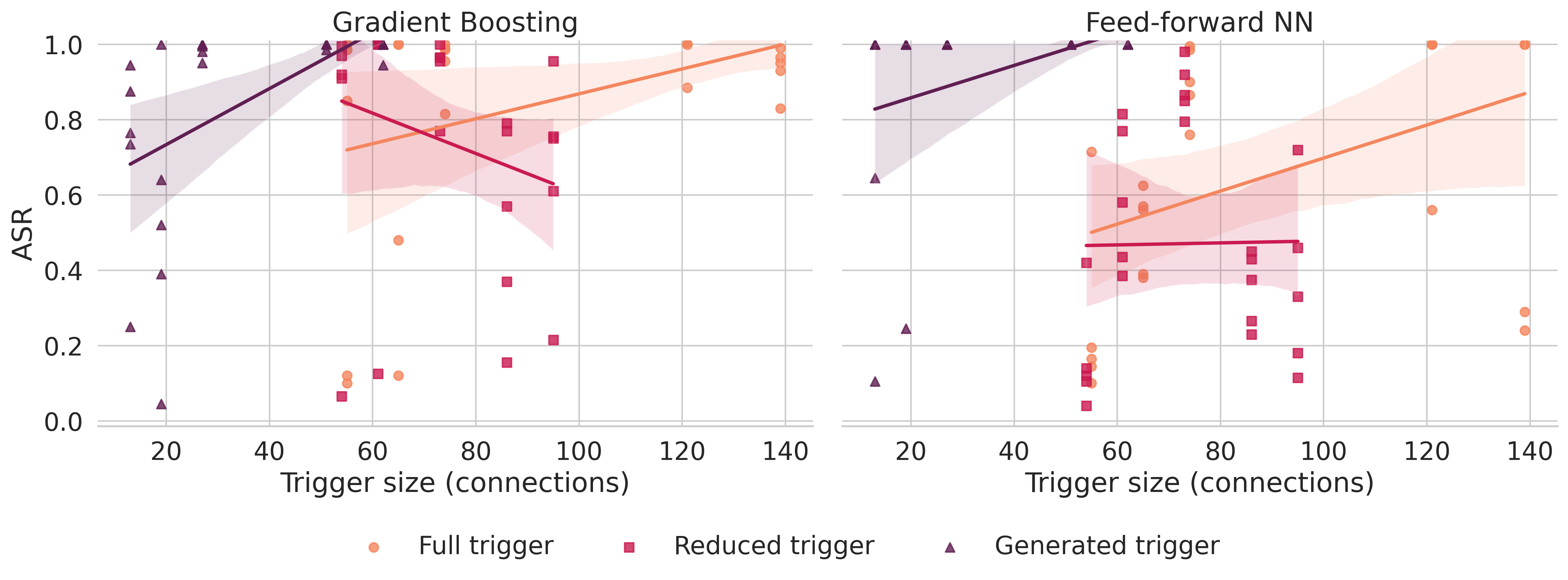}
    \caption{Correlation between the number of connections composing the trigger and the attack success rate (ASR). Each point represents a separate experiment. Curve fitting illustrating the trend is performed using linear regression.}
    \label{fig:neris_entropy_trigger_size}
    \end{subfigure}

    \label{fig:neris_entropy_triggers}
    \caption{Analysis of trigger selection strategy. CTU-13 Neris Botnet scenario, with the Entropy feature selection strategy.}
\vspace{-5pt}
\end{figure*}

\vspace{5pt}
\noindent\textbf{Evaluation of attack success.}
\Cref{fig:entropy_neris} shows the attack success rate as a function of the poisoning percentage for the three different types of triggers: Full, Reduced, and Generated. We observe that all triggers are able to mount effective attacks against the Gradient Boosting classifier, with attack success rates over 0.8 when 0.5\% or more of the training data is poisoned. The Feed-forward Neural Network, is generally more resilient to our attacks: the Full trigger and Reduced trigger deliver an attack success rate of about 0.7 and 0.4, respectively, while the Generative trigger is able to synthesize more effective triggers, which lead to attack success rates over 0.7. 

\Cref{fig:neris_entropy_trigger_size} studies the correlation between trigger size (measured in number of connections) and attack success rate for each type of trigger. Each data  point represented in the figure constitutes a separate experiment, while the regression lines capture the trend (how ASR changes as the trigger size changes). These figures show that the generative method leads to consistently smaller triggers than the other two methods, without sacrificing attack success. 
This result is indicative of the power of generative models in knowledge discovery, and, in our case, their ability to synthesize a small set of realistic log connections that lead to the feature-space prototype. \Cref{fig:neris_entropy_trigger_size} also shows that the size reduction strategy is able to create triggers (Reduced trigger) that are smaller than the Full trigger, but at the expense of the attack success rate.

\vspace{5pt}
\noindent\textbf{Evaluation of attack stealthiness in feature space.}
Next, we evaluate the attack stealthiness in feature space, using the Isolation Forest~\cite{liuIsolationForest2008a} algorithm for anomaly detection.
The objective of this experiment is to see whether a standard technique for anomaly detection can identify and flag the poisoned samples as anomalies.
The anomaly detector is trained on a clean subset of data, which is completely disjoint from the poisoned data points and consists of 10\% of the entire training dataset.

\begin{table}[t]
\centering
\caption{Area under the Precision-Recall Curve and F1 score obtained by performing anomaly detection on the poisoned data with an Isolation Forest model trained on a clean subset of the training data. CTU-13 Neris, at 1\% poisoning rate.}
{\def\arraystretch{1.1}
\small
\begin{tabular}{l|llll}
\hline
\textbf{Strategy}                     & \textbf{Model}        & \textbf{Trigger} & \textbf{PR AUC} & \textbf{$F_1$ score} \\ \hline
\multirow{3}{*}{Entropy} & \multirow{3}{*}{Any} & Full                  & 0.056            & 0.013             \\
                                   &                          & Reduced               & 0.045            & 0.012             \\
                                   &                          & Generated             & 0.078            & 0.018             \\ \hline
\multirow{6}{*}{SHAP}   & \multirow{3}{*}{Gradient Boosting}    & Full                  & 0.099            & 0.015             \\
                                   &                          & Reduced               & 0.070            & 0.013             \\
                                   &                          & Generated             & 0.099            & 0.019             \\ \cline{2-5}
                                   & \multirow{3}{*}{Feed-forward NN}    & Full                  & 0.061            & 0.015             \\
                                   &                          & Reduced               & 0.047            & 0.014             \\
                                   &                          & Generated             & 0.052            & 0.012\\            
\hline
\end{tabular}
}
\label{tab:neris_gb_isof}
\vspace{-5pt}
\end{table}

Table~\ref{tab:neris_gb_isof} presents the anomaly detection results on the poisoned data obtained with each trigger type (Full, Reduced, and Generated). For comparison, we evaluate both the entropy-based and the SHAP-based feature selection strategies used to craft the injected pattern. Since SHAP queries the model to compute feature relevance scores, we present the anomaly detection results separately for a SHAP-guided attack against a Gradient Boosting classifier and against a Feed-forward Neural Network. Across the board, we observe very low Precision-Recall area under the curve (AUC) scores (in the 0.045 -- 0.099 range), as well as very low $F_1$ scores (in the 0.012 -- 0.019 range). These results demonstrate the difficulty of differentiating the poisoned data points from the clean data points, and indicate that the poisoning attacks are highly inconspicuous in feature space.

\vspace{5pt}
\noindent\textbf{Evaluation of attack stealthiness in problem space.}
We also evaluate attack stealthiness in problem space, in terms of how close the poisoned data is to the target class, here represented by the benign class (normal traffic). We leverage the Jensen-Shannon divergence~\cite{Lin1991JensenShannon}, a normalized and symmetrical scoring method for measuring the similarity between two probability distributions, and in particular we use the distance formulation defined as the square root of the divergence, which assumes value of zero for identical distributions. 
We compute the distance for each field in the connection logs (e.g., bytes, port, connection state, etc.), and report the average across all fields.
As a baseline, we compute the average Jensen-Shannon distance between the target class points (benign log connections only) of the training and test datasets, capturing the distribution shift between train and test data.  For the CTU-13 Neris Botnet dataset, we evaluated this reference distance as being \textsc{D\_ref = JS(train, test) = 0.24}.

\begin{figure}[th]
    \centering
    \includegraphics[width=0.9\linewidth]{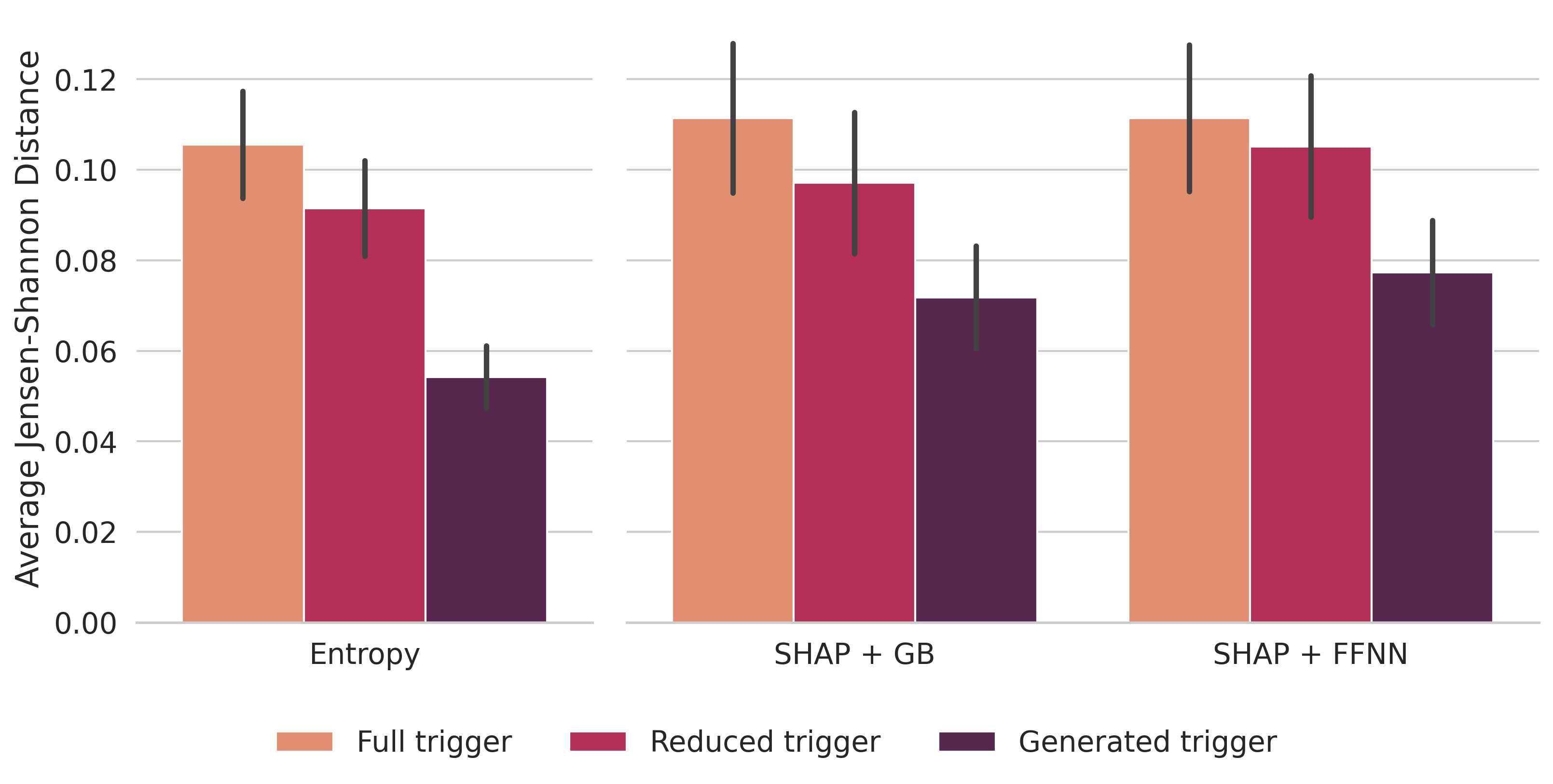}
    \caption{Jensen-Shannon distance between the poisoned and clean training dataset, averaged over all considered \emph{conn.log} fields. For reference, the average JS distance value between the original training data and test data is 0.24. CTU-13 Neris Botnet experiments, at 1\% poisoning rate.}
    \label{fig:neris_avgjs_entropy}
\vspace{-5pt}
\end{figure}

\Cref{fig:neris_avgjs_entropy} shows the Jensen-Shannon distance between the poisoned and clean training dataset for each of the trigger types. The figure illustrates that all three strategies produce stealthy attacks, characterized by average Jensen-Shannon distances that are comfortably lower than \textsc{D\_ref}.
Furthermore, the generative method (Generated trigger) constructs the most inconspicuous triggers, followed by the trigger size reduction method (Reduced trigger).

\subsection{Impact of Feature Representation}
\label{sec:feat_rep}

\begin{table}[t!]
\caption{Results on the CTU-13 Neris Botnet scenario, where the victim model uses an auto-encoder to learn the feature representation. Entropy strategy.}
\centering
\small
\begin{tabular}{l|llllll}
\hline
\textbf{Poison budget} & 0.5\% & 1\%   & 2\%   & 4\%   & 5\%   & 10\%  \\ \hline
\textbf{ASR} & 0.013 & 0.066 & 0.166 & 0.362 & 0.406 & 0.634 \\
\textbf{Stand. dev.} & 0.009 & 0.045 & 0.134 & 0.100 & 0.140 & 0.109 \\ \hline
\textbf{$\Delta F_1$ Test} & 0.002 & 0.003 & 0.005 & 0.009 & 0.011 & 0.007 \\ \hline
\end{tabular}
\label{tab:neris_ae_entropy}
\vspace{-5pt}
\end{table}

The feature representation used by the learning task can strongly influence the attack success. In the next set of experiments, we study feature encodings, which are automatically learned with an auto-encoder architecture. Together with statistical features, encoded features are representative in network traffic classification, and auto-encoder models have been widely adopted for this task by previous works~\cite{mirskyKitsuneEnsembleAutoencoders2018a,yangFeatureExtractionNovelty2021, heDeepFeatureBasedAutoencoderNetwork2021,dangeloNetworkTrafficClassification2021}. 
To generate these features, we first train an auto-encoder model in an unsupervised manner, with the goal of minimizing the reconstruction error. 
Then the encoder portion of the model is run on the same training data to extract the high-level features used to train the feed-forward neural network architecture considered in previous experiments.
Since the auto-encoder requires its inputs to be of a consistent shape, instead of features extracted from 30-second time windows, here the model is provided with an input representation consisting of contiguous blocks of 100 connections. 
Given that features are extracted from connection blocks of a fixed size, we also fix the trigger size to be 50 connections long.
We found this value empirically by experimenting with different trigger sizes, and noticed that smaller ones would lead to unsatisfying attack results.
While the trigger is relatively large compared to the unit block size, it is worth noting that the total number of connections introduced by the attack is still very limited when compared to the size of the the training set.

\Cref{tab:neris_ae_entropy} reports the results of the Entropy strategy when applied in this setup, at different poison percentages, together with its standard deviation across 5 experiments and the average degradation in performance of the victim model on clean data.
Since the auto-encoder was trained in an unsupervised fashion to minimize the reconstruction loss, we expect this training loss to  impact negatively the overall success of the attack.
In fact, we do observe a general reduction of the success rate compared to the simple neural network model, especially for limited poisoning budgets ($\leq 1\%$).
However, if the adversary is allowed to increase the poisoning rate beyond 1\%, we observe that the attack scales nicely with larger poisoning budgets.
At the same time, the $\Delta F_1$ values remain generally low even at larger poison percentages.

\subsection{Other datasets}
\label{sec:other_data}

In the previous sections, we carried out an in-depth evaluation of various attack characteristics and their impact on the attack success. 
In this section, we investigate how generalizable this poisoning approach is by testing it on different datasets and other classification tasks.
We evaluate here a second cybersecurity task on the CIC IDS 2018 dataset, and two application classification scenarios on CIC ISCX 2016.
For all of these case studies, we use the statistical features (see \Cref{tab:features}) and the full trigger strategy.

\begin{figure} 
    \centering
    \includegraphics[width=0.9\linewidth]{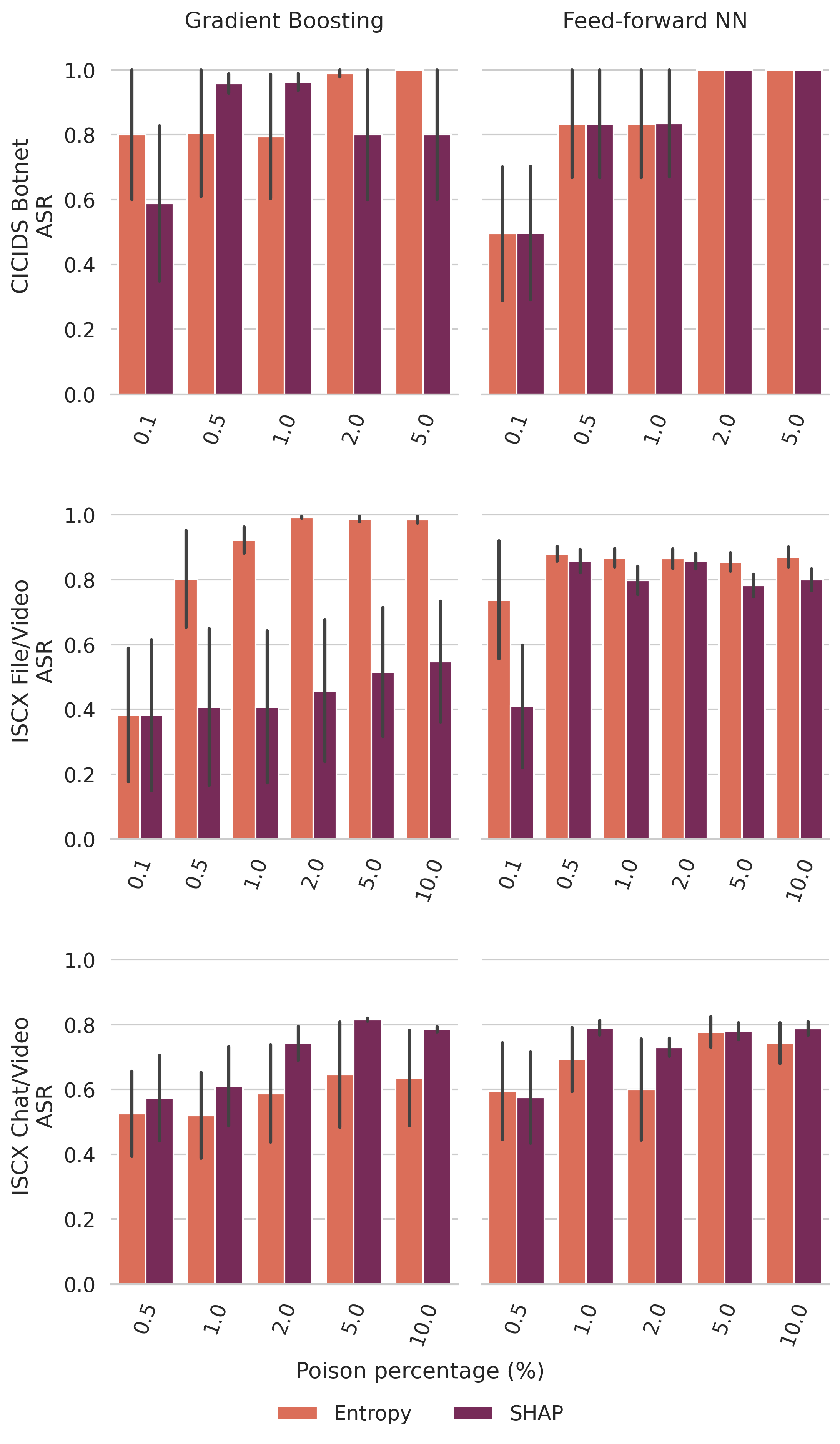}
    \caption{Attack success rate (ASR) on the CIC IDS 2018 Botnet and the CIC ISCX 2016 dataset, full trigger.}
    \vspace{-15pt}
    \label{fig:cic_vs_iscx}
\end{figure}

We report the attack success rate at different poisoning percentages in \Cref{fig:cic_vs_iscx}.
Due to the much smaller size of the ISCX dataset, we test up to slightly larger poison percentage values --- for instance in the Chat/Video scenario, 0.1\% of the training set would amount to a single poisoning point.
In general, we observe similar trends as in previous experiments, with the SHAP and Entropy strategies performing similarly, and achieving significant attack success rates even with very limited poison budgets.

We also evaluated the poisoned model on clean test data, to verify whether the poisoned model is still able to classify clean test data correctly. We obtained very limited reductions in $F_1$ scores: $\Delta F_1$ is between 0.002 and 0.046, with the SHAP strategy resulting in slightly larger shifts than the other feature selection methods.

\section{Discussion and Limitations}
\label{sec:discuss}
Despite our efforts towards the practical feasibility of the attack we propose, poisoning complex data is still a challenging task, and there are some elements that could increase the difficulty of deploying this attack on an arbitrary victim network.
Regarding the problem-space mapping of the triggers, the adversary may experience a situation where two connection events are inter-dependent, due to the internal state of Zeek, but the trigger does not include both of them simultaneously --- this could occur if the connections happen across the border of two time windows.
For instance, inter-dependent connection events may take place in the case of hosts running the FTP protocol.
Documentation on this type of connections for Zeek is quite scarce,
but a dedicated attacker could allocate time and resources to enumerate all possible corner cases and explicitly avoid them during the trigger creation phase.

Another potential source of issues could arise when using the generated trigger approach.
This method leads to a generally good attack success with a small footprint, however, it could in principle generate connections that are not feasible in practice for stateful protocols (TCP). 
There are two possible ways to address this potential issue.
First, given the relentless pace of improvements in generative models, including those targeting tabular data~\cite{xuModelingTabularData2019, bourouReviewTabularData2021}, we expect that the ability of generative models to infer the inter-features constraints that characterize this data modality will increase significantly in the very short term.
In parallel, the adversary could attempt to verify the correctness of the generated connections using a model checker and a formal model of the TCP protocol, and simply reject the non-conforming ones.
Both approaches are exciting avenues for future research, and we leave their in-depth analysis to future work.

Finally, we designed methods to hide the poisoning campaign, and showed that our poisoning points are difficult to identify both in feature space, by using anomaly detection techniques, and in problem space, by analysing the distributional distance of poisoned data.
Defending ML models from backdoor attacks is an open, and extremely complex, research problem. Many of the current proposed solutions are designed to operate in the computer vision domain~\cite{chenDetectingBackdoorAttacks2019}, or on specific model architectures~\cite{liuFinePruningDefendingBackdooring2018,tranSpectralSignaturesBackdoor2018}. In contrast, our attack method generalizes to different model typologies.
Moreover, initial research on defending classifiers from backdoor attacks in the security domain~\cite{hoDataSanitizationApproach2022} highlighted potential trade-offs between robustness and utility (e.g., defenses that rely on data sanitization may mistakenly remove a high number of benign samples in an attempt to prune out potentially poisoned samples).
By releasing new attack strategies, we hope to encourage future research in the challenging direction of defending against backdoor attacks on network traffic.
\section{Conclusions}
With this work we investigated the possibility of carrying out data-only, clean-label, poisoning attacks against network flow classifiers.
We believe this threat model holds substantial significance for the security community, due to its closer alignment with the capabilities exhibited by sophisticated adversaries observed in the wild, and the current best practices in secure ML deployments, in contrast to other prevailing models frequently employed.

The attack strategy we introduce can effectively forge consistent associations between the trigger pattern and the target class even at extremely low poisoning rates (0.1-0.5\% of the training set size). 
This results in notable attack success rates, despite the constrained nature of the attacker. 
While the attack is effective, it has minimal impacts on the victim model's generalization abilities when dealing with clean test data.
Additionally, the detectability of the trigger can be lessened through different strategies to decrease the likelihood of a defender discovering an ongoing poisoning campaign.

Furthermore, we demonstrated that this form of poisoning has a relatively wide applicability for various objectives across different types of classification tasks. The implications of these findings extend our understanding of ML security in practical contexts, and prompt further investigation into effective defense strategies against these refined attack methodologies.

\section*{Acknowledgements}

This research was sponsored by MIT Lincoln Laboratory, the U.S. Army
Combat Capabilities Development Command Army Research Laboratory (DEVCOM ARL) under Cooperative Agreement Number W911NF-13-2-0045, and the Department of Defense
Multidisciplinary Research Program of the University Research Initiative (MURI) under contract W911NF-21-1-0322.

\bibliographystyle{ACM-Reference-Format}
\bibliography{netpois,prj_vpn}


\begin{thebibliography}{90}


\ifx \showCODEN    \undefined \def \showCODEN     #1{\unskip}     \fi
\ifx \showDOI      \undefined \def \showDOI       #1{#1}\fi
\ifx \showISBNx    \undefined \def \showISBNx     #1{\unskip}     \fi
\ifx \showISBNxiii \undefined \def \showISBNxiii  #1{\unskip}     \fi
\ifx \showISSN     \undefined \def \showISSN      #1{\unskip}     \fi
\ifx \showLCCN     \undefined \def \showLCCN      #1{\unskip}     \fi
\ifx \shownote     \undefined \def \shownote      #1{#1}          \fi
\ifx \showarticletitle \undefined \def \showarticletitle #1{#1}   \fi
\ifx \showURL      \undefined \def \showURL       {\relax}        \fi
\providecommand\bibfield[2]{#2}
\providecommand\bibinfo[2]{#2}
\providecommand\natexlab[1]{#1}
\providecommand\showeprint[2][]{arXiv:#2}

\bibitem[Andrew~Marshall and Kumar(2022)]%
        {threat_modeling2022}
\bibfield{author}{\bibinfo{person}{Emre~Kiciman Andrew~Marshall, Jugal~Parikh}
  {and} \bibinfo{person}{Ram Shankar~Siva Kumar}.}
  \bibinfo{year}{2022}\natexlab{}.
\newblock \bibinfo{title}{Threat Modeling {AI/ML} Systems and Dependencies}.
\newblock
  \bibinfo{howpublished}{\url{https://learn.microsoft.com/en-us/security/engineering/threat-modeling-aiml}}.
\newblock


\bibitem[Antonakakis et~al\mbox{.}(2010)]%
        {antonakakis2010building}
\bibfield{author}{\bibinfo{person}{Manos Antonakakis}, \bibinfo{person}{Roberto
  Perdisci}, \bibinfo{person}{David Dagon}, \bibinfo{person}{Wenke Lee}, {and}
  \bibinfo{person}{Nick Feamster}.} \bibinfo{year}{2010}\natexlab{}.
\newblock \showarticletitle{Building a Dynamic Reputation System for {DNS}}. In
  \bibinfo{booktitle}{\emph{Proceedings of the 19th USENIX Conference on
  Security}} (Washington, DC) \emph{(\bibinfo{series}{USENIX Security'10})}.
  \bibinfo{publisher}{USENIX Association}, \bibinfo{address}{USA},
  \bibinfo{pages}{18}.
\newblock
\showISBNx{8887666655554}


\bibitem[Antonakakis et~al\mbox{.}(2011)]%
        {antonakakis11}
\bibfield{author}{\bibinfo{person}{Manos Antonakakis}, \bibinfo{person}{Roberto
  Perdisci}, \bibinfo{person}{Wenke Lee}, \bibinfo{person}{Nikolaos
  Vasiloglou}, {and} \bibinfo{person}{David Dagon}.}
  \bibinfo{year}{2011}\natexlab{}.
\newblock \showarticletitle{Detecting Malware Domains at the Upper {DNS}
  Hierarchy}. In \bibinfo{booktitle}{\emph{Proceedings of the 20th USENIX
  Conference on Security}} (San Francisco, CA)
  \emph{(\bibinfo{series}{SEC'11})}. \bibinfo{publisher}{USENIX Association},
  \bibinfo{address}{USA}, \bibinfo{pages}{27}.
\newblock


\bibitem[Apruzzese et~al\mbox{.}(2019)]%
        {apruzzeseAddressingAdversarialAttacks2019}
\bibfield{author}{\bibinfo{person}{Giovanni Apruzzese},
  \bibinfo{person}{Michele Colajanni}, \bibinfo{person}{Luca Ferretti}, {and}
  \bibinfo{person}{Mirco Marchetti}.} \bibinfo{year}{2019}\natexlab{}.
\newblock \showarticletitle{Addressing {{Adversarial Attacks Against Security
  Systems Based}} on {{Machine Learning}}}. In \bibinfo{booktitle}{\emph{2019
  11th {{International Conference}} on {{Cyber Conflict}} ({{CyCon}})}},
  Vol.~\bibinfo{volume}{900}. \bibinfo{pages}{1--18}.
\newblock
\showISSN{2325-5374}
\urldef\tempurl%
\url{https://doi.org/10.23919/CYCON.2019.8756865}
\showDOI{\tempurl}


\bibitem[Ayub et~al\mbox{.}(2020)]%
        {Ayub2020}
\bibfield{author}{\bibinfo{person}{Md.~Ahsan Ayub}, \bibinfo{person}{William~A.
  Johnson}, \bibinfo{person}{Douglas~A. Talbert}, {and}
  \bibinfo{person}{Ambareen Siraj}.} \bibinfo{year}{2020}\natexlab{}.
\newblock \showarticletitle{Model Evasion Attack on Intrusion Detection Systems
  using Adversarial Machine Learning}. In \bibinfo{booktitle}{\emph{2020 54th
  Annual Conference on Information Sciences and Systems (CISS)}}.
  \bibinfo{pages}{1--6}.
\newblock
\urldef\tempurl%
\url{https://doi.org/10.1109/CISS48834.2020.1570617116}
\showDOI{\tempurl}


\bibitem[Binder et~al\mbox{.}(2016)]%
        {binder2016layer}
\bibfield{author}{\bibinfo{person}{Alexander Binder},
  \bibinfo{person}{Gr{\'e}goire Montavon}, \bibinfo{person}{Sebastian
  Lapuschkin}, \bibinfo{person}{Klaus-Robert M{\"u}ller}, {and}
  \bibinfo{person}{Wojciech Samek}.} \bibinfo{year}{2016}\natexlab{}.
\newblock \showarticletitle{Layer-wise relevance propagation for neural
  networks with local renormalization layers}. In
  \bibinfo{booktitle}{\emph{Artificial Neural Networks and Machine
  Learning--ICANN 2016: 25th International Conference on Artificial Neural
  Networks, Barcelona, Spain, September 6-9, 2016, Proceedings, Part II 25}}.
  Springer, \bibinfo{pages}{63--71}.
\newblock


\bibitem[Bourou et~al\mbox{.}(2021)]%
        {bourouReviewTabularData2021}
\bibfield{author}{\bibinfo{person}{Stavroula Bourou}, \bibinfo{person}{Andreas
  El~Saer}, \bibinfo{person}{Terpsichori-Helen Velivassaki},
  \bibinfo{person}{Artemis Voulkidis}, {and} \bibinfo{person}{Theodore
  Zahariadis}.} \bibinfo{year}{2021}\natexlab{}.
\newblock \showarticletitle{A {{Review}} of {{Tabular Data Synthesis Using
  GANs}} on an {{IDS Dataset}}}.
\newblock \bibinfo{journal}{\emph{Information}} \bibinfo{volume}{12},
  \bibinfo{number}{9} (\bibinfo{date}{Sept.} \bibinfo{year}{2021}),
  \bibinfo{pages}{375}.
\newblock
\showISSN{2078-2489}
\urldef\tempurl%
\url{https://doi.org/10.3390/info12090375}
\showDOI{\tempurl}


\bibitem[Burkhart et~al\mbox{.}(2010)]%
        {Burkhart2010FeatureStatistics}
\bibfield{author}{\bibinfo{person}{Martin Burkhart}, \bibinfo{person}{Mario
  Strasser}, \bibinfo{person}{Dilip Many}, {and} \bibinfo{person}{Xenofontas
  Dimitropoulos}.} \bibinfo{year}{2010}\natexlab{}.
\newblock \showarticletitle{{SEPIA}: {Privacy-Preserving} Aggregation of
  {Multi-Domain} Network Events and Statistics}. In
  \bibinfo{booktitle}{\emph{19th USENIX Security Symposium (USENIX Security
  10)}}. \bibinfo{publisher}{USENIX Association}, \bibinfo{address}{Washington,
  DC}.
\newblock
\urldef\tempurl%
\url{https://www.usenix.org/conference/usenixsecurity10/sepia-privacy-preserving-aggregation-multi-domain-network-events-and}
\showURL{%
\tempurl}


\bibitem[Cao and Gong(2017)]%
        {Cao2017}
\bibfield{author}{\bibinfo{person}{Xiaoyu Cao} {and}
  \bibinfo{person}{Neil~Zhenqiang Gong}.} \bibinfo{year}{2017}\natexlab{}.
\newblock \showarticletitle{Mitigating Evasion Attacks to Deep Neural Networks
  via Region-Based Classification}. In \bibinfo{booktitle}{\emph{Proceedings of
  the 33rd Annual Computer Security Applications Conference}} (Orlando, FL,
  USA) \emph{(\bibinfo{series}{ACSAC '17})}. \bibinfo{publisher}{Association
  for Computing Machinery}, \bibinfo{address}{New York, NY, USA},
  \bibinfo{pages}{278–287}.
\newblock
\showISBNx{9781450353458}
\urldef\tempurl%
\url{https://doi.org/10.1145/3134600.3134606}
\showDOI{\tempurl}


\bibitem[Chen et~al\mbox{.}(2019a)]%
        {chenDetectingBackdoorAttacks2019}
\bibfield{author}{\bibinfo{person}{Bryant Chen}, \bibinfo{person}{Wilka
  Carvalho}, \bibinfo{person}{Nathalie Baracaldo}, \bibinfo{person}{Heiko
  Ludwig}, \bibinfo{person}{Benjamin Edwards}, \bibinfo{person}{Taesung Lee},
  \bibinfo{person}{Ian Molloy}, {and} \bibinfo{person}{Biplav Srivastava}.}
  \bibinfo{year}{2019}\natexlab{a}.
\newblock \showarticletitle{Detecting Backdoor Attacks on Deep Neural Networks
  by Activation Clustering}. In \bibinfo{booktitle}{\emph{Workshop on
  {{Artificial Intelligence Safety}}}}. \bibinfo{publisher}{{CEUR-WS}}.
\newblock
\showISSN{16130073}


\bibitem[Chen et~al\mbox{.}(2019b)]%
        {faketables2019}
\bibfield{author}{\bibinfo{person}{Haipeng Chen}, \bibinfo{person}{Sushil
  Jajodia}, \bibinfo{person}{Jing Liu}, \bibinfo{person}{Noseong Park},
  \bibinfo{person}{Vadim Sokolov}, {and} \bibinfo{person}{V.~S. Subrahmanian}.}
  \bibinfo{year}{2019}\natexlab{b}.
\newblock \showarticletitle{{FakeTables}: Using {GANs} to Generate Functional
  Dependency Preserving Tables with Bounded Real Data}. In
  \bibinfo{booktitle}{\emph{Proceedings of the Twenty-Eighth International
  Joint Conference on Artificial Intelligence, {IJCAI-19}}}.
  \bibinfo{publisher}{International Joint Conferences on Artificial
  Intelligence Organization}, \bibinfo{address}{Macao China},
  \bibinfo{pages}{2074--2080}.
\newblock
\urldef\tempurl%
\url{https://doi.org/10.24963/ijcai.2019/287}
\showDOI{\tempurl}


\bibitem[Chen et~al\mbox{.}(2017)]%
        {Chen2017Backdoor}
\bibfield{author}{\bibinfo{person}{Xinyun Chen}, \bibinfo{person}{Chang Liu},
  \bibinfo{person}{Bo Li}, \bibinfo{person}{Kimberly Lu}, {and}
  \bibinfo{person}{Dawn Song}.} \bibinfo{year}{2017}\natexlab{}.
\newblock \showarticletitle{Targeted Backdoor Attacks on Deep Learning Systems
  Using Data Poisoning}.
\newblock \bibinfo{journal}{\emph{CoRR}}  \bibinfo{volume}{abs/1712.05526}
  (\bibinfo{year}{2017}).
\newblock
\showeprint[arXiv]{1712.05526}
\urldef\tempurl%
\url{http://arxiv.org/abs/1712.05526}
\showURL{%
\tempurl}


\bibitem[Chernikova and Oprea(2022)]%
        {FENCE}
\bibfield{author}{\bibinfo{person}{Alesia Chernikova} {and}
  \bibinfo{person}{Alina Oprea}.} \bibinfo{year}{2022}\natexlab{}.
\newblock \showarticletitle{FENCE: Feasible Evasion Attacks on Neural Networks
  in Constrained Environments}.
\newblock \bibinfo{journal}{\emph{ACM Trans. Priv. Secur.}}
  \bibinfo{volume}{25}, \bibinfo{number}{4}, Article \bibinfo{articleno}{34}
  (\bibinfo{date}{jul} \bibinfo{year}{2022}), \bibinfo{numpages}{34}~pages.
\newblock
\showISSN{2471-2566}
\urldef\tempurl%
\url{https://doi.org/10.1145/3544746}
\showDOI{\tempurl}


\bibitem[D'Angelo and Palmieri(2021)]%
        {dangeloNetworkTrafficClassification2021}
\bibfield{author}{\bibinfo{person}{Gianni D'Angelo} {and}
  \bibinfo{person}{Francesco Palmieri}.} \bibinfo{year}{2021}\natexlab{}.
\newblock \showarticletitle{Network Traffic Classification Using Deep
  Convolutional Recurrent Autoencoder Neural Networks for Spatial\textendash
  Temporal Features Extraction}.
\newblock \bibinfo{journal}{\emph{Journal of Network and Computer
  Applications}}  \bibinfo{volume}{173} (\bibinfo{date}{Jan.}
  \bibinfo{year}{2021}), \bibinfo{pages}{102890}.
\newblock
\showISSN{1084-8045}
\urldef\tempurl%
\url{https://doi.org/10.1016/j.jnca.2020.102890}
\showDOI{\tempurl}


\bibitem[Deleu et~al\mbox{.}(2022)]%
        {deleu2022bayesian}
\bibfield{author}{\bibinfo{person}{Tristan Deleu}, \bibinfo{person}{Ant{\'o}nio
  G{\'o}is}, \bibinfo{person}{Chris~Chinenye Emezue}, \bibinfo{person}{Mansi
  Rankawat}, \bibinfo{person}{Simon Lacoste-Julien}, \bibinfo{person}{Stefan
  Bauer}, {and} \bibinfo{person}{Yoshua Bengio}.}
  \bibinfo{year}{2022}\natexlab{}.
\newblock \showarticletitle{Bayesian Structure Learning with Generative Flow
  Networks}. In \bibinfo{booktitle}{\emph{The 38th Conference on Uncertainty in
  Artificial Intelligence}}.
\newblock


\bibitem[Devarakonda et~al\mbox{.}(2012)]%
        {Devarakonda2012Bayesian}
\bibfield{author}{\bibinfo{person}{Nagaraju Devarakonda},
  \bibinfo{person}{Srinivasulu Pamidi}, \bibinfo{person}{V.~Valli Kumari},
  {and} \bibinfo{person}{A. Govardhan}.} \bibinfo{year}{2012}\natexlab{}.
\newblock \showarticletitle{Intrusion Detection System using Bayesian Network
  and Hidden Markov Model}.
\newblock \bibinfo{journal}{\emph{Procedia Technology}}  \bibinfo{volume}{4}
  (\bibinfo{year}{2012}), \bibinfo{pages}{506--514}.
\newblock
\showISSN{2212-0173}
\urldef\tempurl%
\url{https://doi.org/10.1016/j.protcy.2012.05.081}
\showDOI{\tempurl}
\newblock
\shownote{2nd International Conference on Computer, Communication, Control and
  Information Technology( C3IT-2012) on February 25 - 26, 2012}.


\bibitem[{Draper-Gil} et~al\mbox{.}(2016)]%
        {draper-gilCharacterizationEncryptedVPN2016}
\bibfield{author}{\bibinfo{person}{Gerard {Draper-Gil}},
  \bibinfo{person}{Arash~Habibi Lashkari}, \bibinfo{person}{Mohammad
  Saiful~Islam Mamun}, {and} \bibinfo{person}{Ali A.~Ghorbani}.}
  \bibinfo{year}{2016}\natexlab{}.
\newblock \showarticletitle{Characterization of {{Encrypted}} and {{VPN
  Traffic}} Using {{Time-related Features}}:}. In
  \bibinfo{booktitle}{\emph{Proceedings of the 2nd {{International Conference}}
  on {{Information Systems Security}} and {{Privacy}}}}.
  \bibinfo{publisher}{{SCITEPRESS - Science and and Technology Publications}},
  \bibinfo{address}{{Rome, Italy}}, \bibinfo{pages}{407--414}.
\newblock
\showISBNx{978-989-758-167-0}
\urldef\tempurl%
\url{https://doi.org/10.5220/0005740704070414}
\showDOI{\tempurl}


\bibitem[Engelmann and Lessmann(2021)]%
        {engelmann2021}
\bibfield{author}{\bibinfo{person}{Justin Engelmann} {and}
  \bibinfo{person}{Stefan Lessmann}.} \bibinfo{year}{2021}\natexlab{}.
\newblock \showarticletitle{Conditional {Wasserstein} {GAN}-based Oversampling
  of Tabular Data for Imbalanced Learning}.
\newblock \bibinfo{journal}{\emph{Expert Systems with Applications}}
  \bibinfo{volume}{174} (\bibinfo{date}{01} \bibinfo{year}{2021}),
  \bibinfo{pages}{114582}.
\newblock
\urldef\tempurl%
\url{https://doi.org/10.1016/j.eswa.2021.114582}
\showDOI{\tempurl}


\bibitem[Fan et~al\mbox{.}(2020)]%
        {Fan2020}
\bibfield{author}{\bibinfo{person}{Ju Fan}, \bibinfo{person}{Junyou Chen},
  \bibinfo{person}{Tongyu Liu}, \bibinfo{person}{Yuwei Shen},
  \bibinfo{person}{Guoliang Li}, {and} \bibinfo{person}{Xiaoyong Du}.}
  \bibinfo{year}{2020}\natexlab{}.
\newblock \showarticletitle{Relational data synthesis using generative
  adversarial networks: a design space exploration}.
\newblock \bibinfo{journal}{\emph{Proceedings of the VLDB Endowment}}
  \bibinfo{volume}{13} (\bibinfo{date}{08} \bibinfo{year}{2020}),
  \bibinfo{pages}{1962--1975}.
\newblock
\urldef\tempurl%
\url{https://doi.org/10.14778/3407790.3407802}
\showDOI{\tempurl}


\bibitem[Garc{\'i}a et~al\mbox{.}(2014)]%
        {garciaEmpiricalComparisonBotnet2014}
\bibfield{author}{\bibinfo{person}{S. Garc{\'i}a}, \bibinfo{person}{M. Grill},
  \bibinfo{person}{J. Stiborek}, {and} \bibinfo{person}{A. Zunino}.}
  \bibinfo{year}{2014}\natexlab{}.
\newblock \showarticletitle{An Empirical Comparison of Botnet Detection
  Methods}.
\newblock \bibinfo{journal}{\emph{Computers and Security}}
  \bibinfo{volume}{45} (\bibinfo{date}{Sept.} \bibinfo{year}{2014}),
  \bibinfo{pages}{100--123}.
\newblock
\showISSN{0167-4048}
\urldef\tempurl%
\url{https://doi.org/10.1016/j.cose.2014.05.011}
\showDOI{\tempurl}


\bibitem[Gastwirth(1972)]%
        {Gastwirth72}
\bibfield{author}{\bibinfo{person}{Joseph Gastwirth}.}
  \bibinfo{year}{1972}\natexlab{}.
\newblock \showarticletitle{The Estimation of the {Lorenz Curve and Gini
  Index}}.
\newblock \bibinfo{journal}{\emph{The Review of Economics and Statistics}}
  \bibinfo{volume}{54} (\bibinfo{date}{02} \bibinfo{year}{1972}),
  \bibinfo{pages}{306--16}.
\newblock
\urldef\tempurl%
\url{https://doi.org/10.2307/1937992}
\showDOI{\tempurl}


\bibitem[Grosse et~al\mbox{.}(2023)]%
        {grosseMachineLearningSecurity2023}
\bibfield{author}{\bibinfo{person}{Kathrin Grosse}, \bibinfo{person}{Lukas
  Bieringer}, \bibinfo{person}{Tarek~R. Besold}, \bibinfo{person}{Battista
  Biggio}, {and} \bibinfo{person}{Katharina Krombholz}.}
  \bibinfo{year}{2023}\natexlab{}.
\newblock \showarticletitle{Machine {{Learning Security}} in {{Industry}}: {{A
  Quantitative Survey}}}.
\newblock \bibinfo{journal}{\emph{IEEE Transactions on Information Forensics
  and Security}}  \bibinfo{volume}{18} (\bibinfo{year}{2023}),
  \bibinfo{pages}{1749--1762}.
\newblock
\showISSN{1556-6021}
\urldef\tempurl%
\url{https://doi.org/10.1109/TIFS.2023.3251842}
\showDOI{\tempurl}


\bibitem[Gu et~al\mbox{.}(2019)]%
        {guBadNetsEvaluatingBackdooring2019}
\bibfield{author}{\bibinfo{person}{T. Gu}, \bibinfo{person}{K. Liu},
  \bibinfo{person}{B. {Dolan-Gavitt}}, {and} \bibinfo{person}{S. Garg}.}
  \bibinfo{year}{2019}\natexlab{}.
\newblock \showarticletitle{{{BadNets}}: {{Evaluating Backdooring Attacks}} on
  {{Deep Neural Networks}}}.
\newblock \bibinfo{journal}{\emph{IEEE Access}}  \bibinfo{volume}{7}
  (\bibinfo{year}{2019}), \bibinfo{pages}{47230--47244}.
\newblock
\showISSN{2169-3536}
\urldef\tempurl%
\url{https://doi.org/10.1109/ACCESS.2019.2909068}
\showDOI{\tempurl}


\bibitem[Handley et~al\mbox{.}(2001)]%
        {Handley2001}
\bibfield{author}{\bibinfo{person}{Mark Handley}, \bibinfo{person}{Vern
  Paxson}, {and} \bibinfo{person}{Christian Kreibich}.}
  \bibinfo{year}{2001}\natexlab{}.
\newblock \showarticletitle{Network Intrusion Detection: Evasion, Traffic
  Normalization, and {End-to-End} Protocol Semantics}. In
  \bibinfo{booktitle}{\emph{10th USENIX Security Symposium (USENIX Security
  01)}}. \bibinfo{publisher}{USENIX Association}, \bibinfo{address}{Washington,
  D.C.}
\newblock
\urldef\tempurl%
\url{https://www.usenix.org/conference/10th-usenix-security-symposium/network-intrusion-detection-evasion-traffic-normalization}
\showURL{%
\tempurl}


\bibitem[He et~al\mbox{.}(2021)]%
        {heDeepFeatureBasedAutoencoderNetwork2021}
\bibfield{author}{\bibinfo{person}{Mingshu He}, \bibinfo{person}{Xiaojuan
  Wang}, \bibinfo{person}{Junhua Zhou}, \bibinfo{person}{Yuanyuan Xi},
  \bibinfo{person}{Lei Jin}, {and} \bibinfo{person}{Xinlei Wang}.}
  \bibinfo{year}{2021}\natexlab{}.
\newblock \showarticletitle{Deep-{{Feature-Based Autoencoder Network}} for
  {{Few-Shot Malicious Traffic Detection}}}.
\newblock \bibinfo{journal}{\emph{Security and Communication Networks}}
  \bibinfo{volume}{2021} (\bibinfo{date}{March} \bibinfo{year}{2021}),
  \bibinfo{pages}{e6659022}.
\newblock
\showISSN{1939-0114}
\urldef\tempurl%
\url{https://doi.org/10.1155/2021/6659022}
\showDOI{\tempurl}


\bibitem[Heckerman(2008)]%
        {Heckerman2008}
\bibfield{author}{\bibinfo{person}{David Heckerman}.}
  \bibinfo{year}{2008}\natexlab{}.
\newblock \bibinfo{booktitle}{\emph{A Tutorial on Learning with Bayesian
  Networks}}.
\newblock \bibinfo{publisher}{Springer Berlin Heidelberg},
  \bibinfo{address}{Berlin, Heidelberg}, \bibinfo{pages}{33--82}.
\newblock
\showISBNx{978-3-540-85066-3}
\urldef\tempurl%
\url{https://doi.org/10.1007/978-3-540-85066-3_3}
\showDOI{\tempurl}


\bibitem[Ho et~al\mbox{.}(2022)]%
        {hoDataSanitizationApproach2022}
\bibfield{author}{\bibinfo{person}{Samson Ho}, \bibinfo{person}{Achyut Reddy},
  \bibinfo{person}{Sridhar Venkatesan}, \bibinfo{person}{Rauf Izmailov},
  \bibinfo{person}{Ritu Chadha}, {and} \bibinfo{person}{Alina Oprea}.}
  \bibinfo{year}{2022}\natexlab{}.
\newblock \showarticletitle{Data {{Sanitization Approach}} to {{Mitigate
  Clean-Label Attacks Against Malware Detection Systems}}}. In
  \bibinfo{booktitle}{\emph{{{MILCOM}} 2022 - 2022 {{IEEE Military
  Communications Conference}} ({{MILCOM}})}}. \bibinfo{pages}{993--998}.
\newblock
\showISSN{2155-7586}
\urldef\tempurl%
\url{https://doi.org/10.1109/MILCOM55135.2022.10017768}
\showDOI{\tempurl}


\bibitem[Holland et~al\mbox{.}(2022)]%
        {hollandReproducibleNetworkTraffic2022}
\bibfield{author}{\bibinfo{person}{Jordan Holland}, \bibinfo{person}{Paul
  Schmitt}, \bibinfo{person}{Prateek Mittal}, {and} \bibinfo{person}{Nick
  Feamster}.} \bibinfo{year}{2022}\natexlab{}.
\newblock \bibinfo{title}{Towards {{Reproducible Network Traffic Analysis}}}.
\newblock
\newblock
\showeprint[arxiv]{2203.12410}~[cs]


\bibitem[Holodnak et~al\mbox{.}(2022)]%
        {holodnakBackdoorPoisoningEncrypted2022}
\bibfield{author}{\bibinfo{person}{John~T. Holodnak}, \bibinfo{person}{Olivia
  Brown}, \bibinfo{person}{Jason Matterer}, {and} \bibinfo{person}{Andrew
  Lemke}.} \bibinfo{year}{2022}\natexlab{}.
\newblock \showarticletitle{Backdoor {{Poisoning}} of {{Encrypted Traffic
  Classifiers}}}. In \bibinfo{booktitle}{\emph{2022 {{IEEE International
  Conference}} on {{Data Mining Workshops}} ({{ICDMW}})}}.
  \bibinfo{pages}{577--585}.
\newblock
\showISSN{2375-9259}
\urldef\tempurl%
\url{https://doi.org/10.1109/ICDMW58026.2022.00080}
\showDOI{\tempurl}


\bibitem[IBM(2023)]%
        {ibmqradar}
\bibfield{author}{\bibinfo{person}{IBM}.} \bibinfo{year}{2023}\natexlab{}.
\newblock \bibinfo{title}{{IBM Security QRadar XDR}}.
\newblock
\newblock
\newblock
\shownote{{https://www.ibm.com/qradar}}.


\bibitem[Ingalls(2021)]%
        {xdr_solutions}
\bibfield{author}{\bibinfo{person}{Sam Ingalls}.}
  \bibinfo{year}{2021}\natexlab{}.
\newblock \bibinfo{title}{{Top XDR Security Solutions for 2022}}.
\newblock
  \bibinfo{howpublished}{\url{https://www.esecurityplanet.com/products/xdr-security-solutions/}}.
\newblock


\bibitem[Invernizzi et~al\mbox{.}(2014)]%
        {nazca_ndss2014}
\bibfield{author}{\bibinfo{person}{Luca Invernizzi}, \bibinfo{person}{Sung ju
  Lee}, \bibinfo{person}{Stanislav Miskovic}, \bibinfo{person}{Marco Mellia},
  \bibinfo{person}{Ruben Torres}, \bibinfo{person}{Christopher Kruegel},
  \bibinfo{person}{Sabyasachi Saha}, {and} \bibinfo{person}{Giovanni Vigna}.}
  \bibinfo{year}{2014}\natexlab{}.
\newblock \showarticletitle{{Nazca: Detecting Malware Distribution in
  Large-Scale Networks}}. In \bibinfo{booktitle}{\emph{NDSS}}.
\newblock


\bibitem[Jabbar et~al\mbox{.}(2017)]%
        {Jabbar2017BayesianIDS}
\bibfield{author}{\bibinfo{person}{M.A. Jabbar}, \bibinfo{person}{Rajanikanth
  Aluvalu}, {and} \bibinfo{person}{S.~Sai Satyanarayana~Reddy}.}
  \bibinfo{year}{2017}\natexlab{}.
\newblock \showarticletitle{Intrusion Detection System Using Bayesian Network
  and Feature Subset Selection}. In \bibinfo{booktitle}{\emph{2017 IEEE
  International Conference on Computational Intelligence and Computing Research
  (ICCIC)}}. \bibinfo{pages}{1--5}.
\newblock
\urldef\tempurl%
\url{https://doi.org/10.1109/ICCIC.2017.8524381}
\showDOI{\tempurl}


\bibitem[Jacobs et~al\mbox{.}(2022)]%
        {Jacobs2022Emperor}
\bibfield{author}{\bibinfo{person}{Arthur~S. Jacobs}, \bibinfo{person}{Roman
  Beltiukov}, \bibinfo{person}{Walter Willinger}, \bibinfo{person}{Ronaldo~A.
  Ferreira}, \bibinfo{person}{Arpit Gupta}, {and} \bibinfo{person}{Lisandro~Z.
  Granville}.} \bibinfo{year}{2022}\natexlab{}.
\newblock \showarticletitle{AI/ML for Network Security: The Emperor Has No
  Clothes}. In \bibinfo{booktitle}{\emph{Proceedings of the 2022 ACM SIGSAC
  Conference on Computer and Communications Security}} (Los Angeles, CA, USA)
  \emph{(\bibinfo{series}{CCS '22})}. \bibinfo{publisher}{Association for
  Computing Machinery}, \bibinfo{address}{New York, NY, USA},
  \bibinfo{pages}{1537–1551}.
\newblock
\showISBNx{9781450394505}
\urldef\tempurl%
\url{https://doi.org/10.1145/3548606.3560609}
\showDOI{\tempurl}


\bibitem[Kaur et~al\mbox{.}(2020)]%
        {Kaur2020Bayesian}
\bibfield{author}{\bibinfo{person}{Dhamanpreet Kaur}, \bibinfo{person}{Matthew
  Sobiesk}, \bibinfo{person}{Shubham Patil}, \bibinfo{person}{Jin Liu},
  \bibinfo{person}{Puran Bhagat}, \bibinfo{person}{Amar Gupta}, {and}
  \bibinfo{person}{Natasha Markuzon}.} \bibinfo{year}{2020}\natexlab{}.
\newblock \showarticletitle{Application of Bayesian networks to generate
  synthetic health data}.
\newblock \bibinfo{journal}{\emph{Journal of the American Medical Informatics
  Association}}  \bibinfo{volume}{28} (\bibinfo{date}{12}
  \bibinfo{year}{2020}).
\newblock
\urldef\tempurl%
\url{https://doi.org/10.1093/jamia/ocaa303}
\showDOI{\tempurl}


\bibitem[Kitson et~al\mbox{.}(2023)]%
        {Kitson2023Bayesian}
\bibfield{author}{\bibinfo{person}{Neville~Kenneth Kitson},
  \bibinfo{person}{Anthony~C. Constantinou}, \bibinfo{person}{Zhigao Guo},
  \bibinfo{person}{Yang Liu}, {and} \bibinfo{person}{Kiattikun Chobtham}.}
  \bibinfo{year}{2023}\natexlab{}.
\newblock \showarticletitle{A survey of Bayesian Network structure learning}.
\newblock \bibinfo{journal}{\emph{Artificial Intelligence Review}}
  (\bibinfo{year}{2023}).
\newblock
\showISBNx{1573-7462}
\urldef\tempurl%
\url{https://doi.org/10.1007/s10462-022-10351-w}
\showDOI{\tempurl}


\bibitem[Koller and Sahami(1996)]%
        {Koller96}
\bibfield{author}{\bibinfo{person}{Daphne Koller} {and} \bibinfo{person}{Mehran
  Sahami}.} \bibinfo{year}{1996}\natexlab{}.
\newblock \showarticletitle{Toward Optimal Feature Selection}. In
  \bibinfo{booktitle}{\emph{Proceedings of the Thirteenth International
  Conference on International Conference on Machine Learning}} (Bari, Italy)
  \emph{(\bibinfo{series}{ICML'96})}. \bibinfo{publisher}{Morgan Kaufmann
  Publishers Inc.}, \bibinfo{address}{San Francisco, CA, USA},
  \bibinfo{pages}{284–292}.
\newblock
\showISBNx{1558604197}


\bibitem[Kotelnikov et~al\mbox{.}(2022)]%
        {kotelnikovTabDDPMModellingTabular2022}
\bibfield{author}{\bibinfo{person}{Akim Kotelnikov}, \bibinfo{person}{Dmitry
  Baranchuk}, \bibinfo{person}{Ivan Rubachev}, {and} \bibinfo{person}{Artem
  Babenko}.} \bibinfo{year}{2022}\natexlab{}.
\newblock \bibinfo{title}{{{TabDDPM}}: {{Modelling Tabular Data}} with
  {{Diffusion Models}}}.
\newblock
\newblock
\urldef\tempurl%
\url{https://doi.org/10.48550/arXiv.2209.15421}
\showDOI{\tempurl}
\showeprint[arxiv]{2209.15421}~[cs]


\bibitem[Lee and Lee(2006)]%
        {LEE2006}
\bibfield{author}{\bibinfo{person}{Changki Lee} {and}
  \bibinfo{person}{Gary~Geunbae Lee}.} \bibinfo{year}{2006}\natexlab{}.
\newblock \showarticletitle{Information gain and divergence-based feature
  selection for machine learning-based text categorization}.
\newblock \bibinfo{journal}{\emph{Information Processing \& Management}}
  \bibinfo{volume}{42}, \bibinfo{number}{1} (\bibinfo{year}{2006}),
  \bibinfo{pages}{155--165}.
\newblock
\showISSN{0306-4573}
\urldef\tempurl%
\url{https://doi.org/10.1016/j.ipm.2004.08.006}
\showDOI{\tempurl}
\newblock
\shownote{Formal Methods for Information Retrieval}.


\bibitem[Li et~al\mbox{.}(2018a)]%
        {liChronicPoisoningMachine2018a}
\bibfield{author}{\bibinfo{person}{Pan Li}, \bibinfo{person}{Qiang Liu},
  \bibinfo{person}{Wentao Zhao}, \bibinfo{person}{Dongxu Wang}, {and}
  \bibinfo{person}{Siqi Wang}.} \bibinfo{year}{2018}\natexlab{a}.
\newblock \showarticletitle{Chronic {{Poisoning}} against {{Machine Learning
  Based IDSs Using Edge Pattern Detection}}}. In \bibinfo{booktitle}{\emph{2018
  {{IEEE International Conference}} on {{Communications}} ({{ICC}})}}.
  \bibinfo{pages}{1--7}.
\newblock
\showISSN{1938-1883}
\urldef\tempurl%
\url{https://doi.org/10.1109/ICC.2018.8422328}
\showDOI{\tempurl}


\bibitem[Li et~al\mbox{.}(2018b)]%
        {Li2018EPD}
\bibfield{author}{\bibinfo{person}{Pan Li}, \bibinfo{person}{Qiang Liu},
  \bibinfo{person}{Wentao Zhao}, \bibinfo{person}{Dongxu Wang}, {and}
  \bibinfo{person}{Siqi Wang}.} \bibinfo{year}{2018}\natexlab{b}.
\newblock \showarticletitle{Chronic Poisoning against Machine Learning Based
  IDSs Using Edge Pattern Detection}. In \bibinfo{booktitle}{\emph{2018 IEEE
  International Conference on Communications (ICC)}}. \bibinfo{pages}{1--7}.
\newblock
\urldef\tempurl%
\url{https://doi.org/10.1109/ICC.2018.8422328}
\showDOI{\tempurl}


\bibitem[Lin(1991)]%
        {Lin1991JensenShannon}
\bibfield{author}{\bibinfo{person}{J. Lin}.} \bibinfo{year}{1991}\natexlab{}.
\newblock \showarticletitle{Divergence measures based on the Shannon entropy}.
\newblock \bibinfo{journal}{\emph{IEEE Transactions on Information Theory}}
  \bibinfo{volume}{37}, \bibinfo{number}{1} (\bibinfo{year}{1991}),
  \bibinfo{pages}{145--151}.
\newblock
\urldef\tempurl%
\url{https://doi.org/10.1109/18.61115}
\showDOI{\tempurl}


\bibitem[Linardatos et~al\mbox{.}(2021)]%
        {Linardatos2021}
\bibfield{author}{\bibinfo{person}{Pantelis Linardatos},
  \bibinfo{person}{Vasilis Papastefanopoulos}, {and} \bibinfo{person}{Sotiris
  Kotsiantis}.} \bibinfo{year}{2021}\natexlab{}.
\newblock \showarticletitle{Explainable AI: A Review of Machine Learning
  Interpretability Methods}.
\newblock \bibinfo{journal}{\emph{Entropy}} \bibinfo{volume}{23},
  \bibinfo{number}{1} (\bibinfo{year}{2021}).
\newblock
\showISSN{1099-4300}
\urldef\tempurl%
\url{https://doi.org/10.3390/e23010018}
\showDOI{\tempurl}


\bibitem[Liu et~al\mbox{.}(2008)]%
        {liuIsolationForest2008a}
\bibfield{author}{\bibinfo{person}{Fei~Tony Liu}, \bibinfo{person}{Kai~Ming
  Ting}, {and} \bibinfo{person}{Zhi-Hua Zhou}.}
  \bibinfo{year}{2008}\natexlab{}.
\newblock \showarticletitle{Isolation {{Forest}}}. In
  \bibinfo{booktitle}{\emph{2008 {{Eighth IEEE International Conference}} on
  {{Data Mining}}}}. \bibinfo{publisher}{{IEEE}}, \bibinfo{address}{{Pisa,
  Italy}}, \bibinfo{pages}{413--422}.
\newblock
\showISBNx{978-0-7695-3502-9}
\urldef\tempurl%
\url{https://doi.org/10.1109/ICDM.2008.17}
\showDOI{\tempurl}


\bibitem[Liu et~al\mbox{.}(2018a)]%
        {liuFinePruningDefendingBackdooring2018}
\bibfield{author}{\bibinfo{person}{Kang Liu}, \bibinfo{person}{Brendan
  {Dolan-Gavitt}}, {and} \bibinfo{person}{Siddharth Garg}.}
  \bibinfo{year}{2018}\natexlab{a}.
\newblock \showarticletitle{Fine-{{Pruning}}: {{Defending Against Backdooring
  Attacks}} on {{Deep Neural Networks}}}. In \bibinfo{booktitle}{\emph{Research
  in {{Attacks}}, {{Intrusions}}, and {{Defenses}}}}
  \emph{(\bibinfo{series}{Lecture {{Notes}} in {{Computer Science}}})},
  \bibfield{editor}{\bibinfo{person}{Michael Bailey}, \bibinfo{person}{Thorsten
  Holz}, \bibinfo{person}{Manolis Stamatogiannakis}, {and}
  \bibinfo{person}{Sotiris Ioannidis}} (Eds.). \bibinfo{publisher}{{Springer
  International Publishing}}, \bibinfo{address}{{Cham}},
  \bibinfo{pages}{273--294}.
\newblock
\showISBNx{978-3-030-00470-5}
\urldef\tempurl%
\url{https://doi.org/10.1007/978-3-030-00470-5_13}
\showDOI{\tempurl}


\bibitem[Liu et~al\mbox{.}(2018b)]%
        {Liu2018Trojaning}
\bibfield{author}{\bibinfo{person}{Yingqi Liu}, \bibinfo{person}{Shiqing Ma},
  \bibinfo{person}{Yousra Aafer}, \bibinfo{person}{Wen{-}Chuan Lee},
  \bibinfo{person}{Juan Zhai}, \bibinfo{person}{Weihang Wang}, {and}
  \bibinfo{person}{Xiangyu Zhang}.} \bibinfo{year}{2018}\natexlab{b}.
\newblock \showarticletitle{Trojaning Attack on Neural Networks}. In
  \bibinfo{booktitle}{\emph{25th Annual Network and Distributed System Security
  Symposium, {NDSS} 2018, San Diego, California, USA, February 18-21, 2018}}.
  \bibinfo{publisher}{The Internet Society}.
\newblock
\urldef\tempurl%
\url{http://wp.internetsociety.org/ndss/wp-content/uploads/sites/25/2018/02/ndss2018\_03A-5\_Liu\_paper.pdf}
\showURL{%
\tempurl}


\bibitem[Lundberg et~al\mbox{.}(2020)]%
        {Lundberg2020}
\bibfield{author}{\bibinfo{person}{Scott~M. Lundberg}, \bibinfo{person}{Gabriel
  Erion}, \bibinfo{person}{Hugh Chen}, \bibinfo{person}{Alex DeGrave},
  \bibinfo{person}{Jordan~M. Prutkin}, \bibinfo{person}{Bala Nair},
  \bibinfo{person}{Ronit Katz}, \bibinfo{person}{Jonathan Himmelfarb},
  \bibinfo{person}{Nisha Bansal}, {and} \bibinfo{person}{Su-In Lee}.}
  \bibinfo{year}{2020}\natexlab{}.
\newblock \showarticletitle{From local explanations to global understanding
  with explainable AI for trees}.
\newblock \bibinfo{journal}{\emph{Nature Machine Intelligence}}
  \bibinfo{volume}{2}, \bibinfo{number}{1} (\bibinfo{year}{2020}),
  \bibinfo{pages}{56--67}.
\newblock
\showISBNx{2522-5839}
\urldef\tempurl%
\url{https://doi.org/10.1038/s42256-019-0138-9}
\showDOI{\tempurl}


\bibitem[Lundberg and Lee(2017)]%
        {SHAP}
\bibfield{author}{\bibinfo{person}{Scott~M Lundberg} {and}
  \bibinfo{person}{Su-In Lee}.} \bibinfo{year}{2017}\natexlab{}.
\newblock \showarticletitle{A Unified Approach to Interpreting Model
  Predictions}. In \bibinfo{booktitle}{\emph{Advances in Neural Information
  Processing Systems}}, \bibfield{editor}{\bibinfo{person}{I.~Guyon},
  \bibinfo{person}{U.~Von Luxburg}, \bibinfo{person}{S.~Bengio},
  \bibinfo{person}{H.~Wallach}, \bibinfo{person}{R.~Fergus},
  \bibinfo{person}{S.~Vishwanathan}, {and} \bibinfo{person}{R.~Garnett}}
  (Eds.), Vol.~\bibinfo{volume}{30}. \bibinfo{publisher}{Curran Associates,
  Inc.}
\newblock
\urldef\tempurl%
\url{https://proceedings.neurips.cc/paper_files/paper/2017/file/8a20a8621978632d76c43dfd28b67767-Paper.pdf}
\showURL{%
\tempurl}


\bibitem[MalwareGuard FireEye(2020)]%
        {fireeye2020Jun}
MalwareGuard FireEye \bibinfo{year}{2020}\natexlab{}.
\newblock \bibinfo{title}{{MalwareGuard: FireEye{'}s Machine Learning Model to
  Detect and Prevent Malware}}.
\newblock
\newblock
\newblock
\shownote{{https://www.fireeye.com/blog/products-and-services/2018/07/malwareguard-fireeye-machine-learning-model-to-detect-and-prevent-malware.html}}.


\bibitem[Microsoft(2021)]%
        {microsoftdefender}
\bibfield{author}{\bibinfo{person}{Microsoft}.}
  \bibinfo{year}{2021}\natexlab{}.
\newblock \bibinfo{title}{{Microsoft Defender for Endpoint {$\vert$} Microsoft
  Security}}.
\newblock
\newblock
\urldef\tempurl%
\url{https://www.microsoft.com/en-us/security/business/threat-protection/endpoint-defender}
\showURL{%
\tempurl}


\bibitem[Mirsky et~al\mbox{.}(2018)]%
        {mirskyKitsuneEnsembleAutoencoders2018a}
\bibfield{author}{\bibinfo{person}{Yisroel Mirsky}, \bibinfo{person}{Tomer
  Doitshman}, \bibinfo{person}{Yuval Elovici}, {and} \bibinfo{person}{Asaf
  Shabtai}.} \bibinfo{year}{2018}\natexlab{}.
\newblock \showarticletitle{Kitsune: {{An Ensemble}} of {{Autoencoders}} for
  {{Online Network Intrusion Detection}}}. In
  \bibinfo{booktitle}{\emph{Proceedings 2018 {{Network}} and {{Distributed
  System Security Symposium}}}}. \bibinfo{publisher}{{Internet Society}},
  \bibinfo{address}{{San Diego, CA}}.
\newblock
\showISBNx{978-1-891562-49-5}
\urldef\tempurl%
\url{https://doi.org/10.14722/ndss.2018.23204}
\showDOI{\tempurl}


\bibitem[Moore et~al\mbox{.}(2005)]%
        {mooreDiscriminatorsUseFlowbased2005}
\bibfield{author}{\bibinfo{person}{Andrew Moore}, \bibinfo{person}{Denis Zuev},
  {and} \bibinfo{person}{Michael Crogan}.} \bibinfo{year}{2005}\natexlab{}.
\newblock \bibinfo{booktitle}{\emph{Discriminators for Use in Flow-Based
  Classification}}.
\newblock \bibinfo{type}{{T}echnical {R}eport}. \bibinfo{institution}{{Queen
  Mary and Westfield College, Department of Computer Science}}.
\newblock


\bibitem[Mukherjee et~al\mbox{.}(1994)]%
        {mukherjeeNetworkIntrusionDetection1994a}
\bibfield{author}{\bibinfo{person}{B. Mukherjee}, \bibinfo{person}{L.T.
  Heberlein}, {and} \bibinfo{person}{K.N. Levitt}.}
  \bibinfo{year}{1994}\natexlab{}.
\newblock \showarticletitle{Network Intrusion Detection}.
\newblock \bibinfo{journal}{\emph{IEEE Network}} \bibinfo{volume}{8},
  \bibinfo{number}{3} (\bibinfo{date}{May} \bibinfo{year}{1994}),
  \bibinfo{pages}{26--41}.
\newblock
\showISSN{1558-156X}
\urldef\tempurl%
\url{https://doi.org/10.1109/65.283931}
\showDOI{\tempurl}


\bibitem[Nelms et~al\mbox{.}(2013)]%
        {nelms2013execscent}
\bibfield{author}{\bibinfo{person}{Terry Nelms}, \bibinfo{person}{Roberto
  Perdisci}, {and} \bibinfo{person}{Mustaque Ahamad}.}
  \bibinfo{year}{2013}\natexlab{}.
\newblock \showarticletitle{{ExecScent: Mining for New C\&C Domains in Live
  Networks with Adaptive Control Protocol Templates}}. In
  \bibinfo{booktitle}{\emph{Proceedings of the 22nd USENIX Conf. on Security}}.
  \bibinfo{publisher}{USENIX Association}, \bibinfo{address}{USA},
  \bibinfo{pages}{589–604}.
\newblock


\bibitem[Nelson et~al\mbox{.}(2008)]%
        {Nelson2008}
\bibfield{author}{\bibinfo{person}{Blaine Nelson}, \bibinfo{person}{Marco
  Barreno}, \bibinfo{person}{Fuching~Jack Chi}, \bibinfo{person}{Anthony~D.
  Joseph}, \bibinfo{person}{Benjamin I.~P. Rubinstein}, \bibinfo{person}{Udam
  Saini}, \bibinfo{person}{Charles Sutton}, \bibinfo{person}{J.~D. Tygar},
  {and} \bibinfo{person}{Kai Xia}.} \bibinfo{year}{2008}\natexlab{}.
\newblock \showarticletitle{Exploiting Machine Learning to Subvert Your Spam
  Filter}. In \bibinfo{booktitle}{\emph{Proceedings of the 1st Usenix Workshop
  on Large-Scale Exploits and Emergent Threats}} (San Francisco, California)
  \emph{(\bibinfo{series}{LEET'08})}. \bibinfo{publisher}{USENIX Association},
  \bibinfo{address}{USA}, Article \bibinfo{articleno}{7},
  \bibinfo{numpages}{9}~pages.
\newblock


\bibitem[Newsome et~al\mbox{.}(2006)]%
        {Newsome2006Paragraph}
\bibfield{author}{\bibinfo{person}{James Newsome}, \bibinfo{person}{Brad Karp},
  {and} \bibinfo{person}{Dawn Song}.} \bibinfo{year}{2006}\natexlab{}.
\newblock \showarticletitle{Paragraph: Thwarting Signature Learning by Training
  Maliciously}. In \bibinfo{booktitle}{\emph{Recent Advances in Intrusion
  Detection}}, \bibfield{editor}{\bibinfo{person}{Diego Zamboni} {and}
  \bibinfo{person}{Christopher Kruegel}} (Eds.). \bibinfo{publisher}{Springer
  Berlin Heidelberg}, \bibinfo{address}{Berlin, Heidelberg},
  \bibinfo{pages}{81--105}.
\newblock
\showISBNx{978-3-540-39725-0}


\bibitem[Ning et~al\mbox{.}(2022)]%
        {ningTrojanFlowNeuralBackdoor2022}
\bibfield{author}{\bibinfo{person}{Rui Ning}, \bibinfo{person}{Chunsheng Xin},
  {and} \bibinfo{person}{Hongyi Wu}.} \bibinfo{year}{2022}\natexlab{}.
\newblock \showarticletitle{{{TrojanFlow}}: {{A Neural Backdoor Attack}} to
  {{Deep Learning-based Network Traffic Classifiers}}}. In
  \bibinfo{booktitle}{\emph{{{IEEE INFOCOM}} 2022 - {{IEEE Conference}} on
  {{Computer Communications}}}}. \bibinfo{pages}{1429--1438}.
\newblock
\showISSN{2641-9874}
\urldef\tempurl%
\url{https://doi.org/10.1109/INFOCOM48880.2022.9796878}
\showDOI{\tempurl}


\bibitem[Ongun et~al\mbox{.}(2022)]%
        {CELEST}
\bibfield{author}{\bibinfo{person}{Talha Ongun}, \bibinfo{person}{Simona
  Boboila}, \bibinfo{person}{Alina Oprea}, \bibinfo{person}{Tina
  Eliassi{-}Rad}, \bibinfo{person}{Jason Hiser}, {and} \bibinfo{person}{Jack~W.
  Davidson}.} \bibinfo{year}{2022}\natexlab{}.
\newblock \showarticletitle{{CELEST}: Federated Learning for Globally
  Coordinated Threat Detection}.
\newblock \bibinfo{journal}{\emph{CoRR}}  \bibinfo{volume}{abs/2205.11459}
  (\bibinfo{year}{2022}).
\newblock
\urldef\tempurl%
\url{https://doi.org/10.48550/arXiv.2205.11459}
\showDOI{\tempurl}
\showeprint[arXiv]{2205.11459}


\bibitem[Ongun et~al\mbox{.}(2019)]%
        {ongunDesigningMachineLearning2019}
\bibfield{author}{\bibinfo{person}{Talha Ongun}, \bibinfo{person}{Timothy
  Sakharaov}, \bibinfo{person}{Simona Boboila}, \bibinfo{person}{Alina Oprea},
  {and} \bibinfo{person}{Tina {Eliassi-Rad}}.} \bibinfo{year}{2019}\natexlab{}.
\newblock \bibinfo{title}{On {{Designing Machine Learning Models}} for
  {{Malicious Network Traffic Classification}}}.
\newblock
\newblock
\showeprint[arxiv]{1907.04846}~[cs, stat]


\bibitem[Ongun et~al\mbox{.}(2021)]%
        {ongunPORTFILERPortLevelNetwork2021}
\bibfield{author}{\bibinfo{person}{Talha Ongun}, \bibinfo{person}{Oliver
  Spohngellert}, \bibinfo{person}{Benjamin Miller}, \bibinfo{person}{Simona
  Boboila}, \bibinfo{person}{Alina Oprea}, \bibinfo{person}{Tina
  {Eliassi-Rad}}, \bibinfo{person}{Jason Hiser}, \bibinfo{person}{Alastair
  Nottingham}, \bibinfo{person}{Jack Davidson}, {and} \bibinfo{person}{Malathi
  Veeraraghavan}.} \bibinfo{year}{2021}\natexlab{}.
\newblock \showarticletitle{{{PORTFILER}}: {{Port-Level Network Profiling}} for
  {{Self-Propagating Malware Detection}}}. In \bibinfo{booktitle}{\emph{2021
  {{IEEE Conference}} on {{Communications}} and {{Network Security}}
  ({{CNS}})}}. \bibinfo{pages}{182--190}.
\newblock
\urldef\tempurl%
\url{https://doi.org/10.1109/CNS53000.2021.9705045}
\showDOI{\tempurl}


\bibitem[Oprea et~al\mbox{.}(2018)]%
        {MADE}
\bibfield{author}{\bibinfo{person}{Alina Oprea}, \bibinfo{person}{Zhou Li},
  \bibinfo{person}{Robin Norris}, {and} \bibinfo{person}{Kevin Bowers}.}
  \bibinfo{year}{2018}\natexlab{}.
\newblock \showarticletitle{{MADE}: Security Analytics for Enterprise Threat
  Detection}. In \bibinfo{booktitle}{\emph{Proceedings of Annual Computer
  Security Applications Conference}} \emph{(\bibinfo{series}{ACSAC})}.
\newblock
\urldef\tempurl%
\url{https://doi.org/10.1145/3274694.3274710}
\showDOI{\tempurl}


\bibitem[Oprea and Vassilev(2023)]%
        {opreaAdversarialMachineLearning2023}
\bibfield{author}{\bibinfo{person}{Alina Oprea} {and} \bibinfo{person}{Apostol
  Vassilev}.} \bibinfo{year}{2023}\natexlab{}.
\newblock \bibinfo{booktitle}{\emph{Adversarial {{Machine Learning}}: {{A
  Taxonomy}} and {{Terminology}} of {{Attacks}} and {{Mitigations}}
  ({{Draft}})}}.
\newblock \bibinfo{type}{{T}echnical {R}eport} NIST AI 100-2e2023 ipd.
  \bibinfo{institution}{{National Institute of Standards and Technology}}.
\newblock


\bibitem[Papadopoulos et~al\mbox{.}(2021a)]%
        {papadopoulosLaunchingAdversarialAttacks2021}
\bibfield{author}{\bibinfo{person}{Pavlos Papadopoulos},
  \bibinfo{person}{Oliver {Thornewill von Essen}}, \bibinfo{person}{Nikolaos
  Pitropakis}, \bibinfo{person}{Christos Chrysoulas}, \bibinfo{person}{Alexios
  Mylonas}, {and} \bibinfo{person}{William~J. Buchanan}.}
  \bibinfo{year}{2021}\natexlab{a}.
\newblock \showarticletitle{Launching {{Adversarial Attacks}} against {{Network
  Intrusion Detection Systems}} for {{IoT}}}.
\newblock \bibinfo{journal}{\emph{Journal of Cybersecurity and Privacy}}
  \bibinfo{volume}{1}, \bibinfo{number}{2} (\bibinfo{date}{June}
  \bibinfo{year}{2021}), \bibinfo{pages}{252--273}.
\newblock
\showISSN{2624-800X}
\urldef\tempurl%
\url{https://doi.org/10.3390/jcp1020014}
\showDOI{\tempurl}


\bibitem[Papadopoulos et~al\mbox{.}(2021b)]%
        {Papadopoulos2021}
\bibfield{author}{\bibinfo{person}{Pavlos Papadopoulos},
  \bibinfo{person}{Oliver Thornewill~von Essen}, \bibinfo{person}{Nikolaos
  Pitropakis}, \bibinfo{person}{Christos Chrysoulas}, \bibinfo{person}{Alexios
  Mylonas}, {and} \bibinfo{person}{William~J. Buchanan}.}
  \bibinfo{year}{2021}\natexlab{b}.
\newblock \showarticletitle{{Launching Adversarial Attacks against Network
  Intrusion Detection Systems for IoT}}.
\newblock \bibinfo{journal}{\emph{Journal of Cybersecurity and Privacy}}
  \bibinfo{volume}{1}, \bibinfo{number}{2} (\bibinfo{year}{2021}),
  \bibinfo{pages}{252--273}.
\newblock
\showISSN{2624-800X}
\urldef\tempurl%
\url{https://doi.org/10.3390/jcp1020014}
\showDOI{\tempurl}


\bibitem[Perdisci et~al\mbox{.}(2006)]%
        {Perdisci2006Worm}
\bibfield{author}{\bibinfo{person}{R. Perdisci}, \bibinfo{person}{M. Sharif},
  \bibinfo{person}{P. Fogla}, \bibinfo{person}{W. Lee}, {and}
  \bibinfo{person}{D. Dagon}.} \bibinfo{year}{2006}\natexlab{}.
\newblock \showarticletitle{Misleading Worm Signature Generators Using
  Deliberate Noise Injection}. In \bibinfo{booktitle}{\emph{2012 IEEE Symposium
  on Security and Privacy}}. \bibinfo{publisher}{IEEE Computer Society},
  \bibinfo{address}{Los Alamitos, CA, USA}, \bibinfo{pages}{17--31}.
\newblock
\showISSN{1081-6011}
\urldef\tempurl%
\url{https://doi.org/10.1109/SP.2006.26}
\showDOI{\tempurl}


\bibitem[Philipp et~al\mbox{.}(2021)]%
        {Philipp2021MLaas}
\bibfield{author}{\bibinfo{person}{Robert Philipp}, \bibinfo{person}{Andreas
  Mladenow}, \bibinfo{person}{Christine Strauss}, {and}
  \bibinfo{person}{Alexander V\"{o}lz}.} \bibinfo{year}{2021}\natexlab{}.
\newblock \showarticletitle{Machine Learning as a Service: Challenges in
  Research and Applications}. In \bibinfo{booktitle}{\emph{Proceedings of the
  22nd International Conference on Information Integration and Web-Based
  Applications \& Services}} (Chiang Mai, Thailand)
  \emph{(\bibinfo{series}{iiWAS '20})}. \bibinfo{publisher}{Association for
  Computing Machinery}, \bibinfo{address}{New York, NY, USA},
  \bibinfo{pages}{396–406}.
\newblock
\showISBNx{9781450389228}
\urldef\tempurl%
\url{https://doi.org/10.1145/3428757.3429152}
\showDOI{\tempurl}


\bibitem[Pierazzi et~al\mbox{.}(2020)]%
        {pierazziIntriguingPropertiesAdversarial2020}
\bibfield{author}{\bibinfo{person}{Fabio Pierazzi}, \bibinfo{person}{Feargus
  Pendlebury}, \bibinfo{person}{Jacopo Cortellazzi}, {and}
  \bibinfo{person}{Lorenzo Cavallaro}.} \bibinfo{year}{2020}\natexlab{}.
\newblock \showarticletitle{Intriguing {{Properties}} of {{Adversarial ML
  Attacks}} in the {{Problem Space}}}. In \bibinfo{booktitle}{\emph{2020 {{IEEE
  Symposium}} on {{Security}} and {{Privacy}} ({{SP}})}}.
  \bibinfo{publisher}{{IEEE}}, \bibinfo{address}{{San Francisco, CA, USA}},
  \bibinfo{pages}{1332--1349}.
\newblock
\showISBNx{978-1-72813-497-0}
\urldef\tempurl%
\url{https://doi.org/10.1109/SP40000.2020.00073}
\showDOI{\tempurl}


\bibitem[Rahbarinia et~al\mbox{.}(2015)]%
        {segugio}
\bibfield{author}{\bibinfo{person}{Babak Rahbarinia}, \bibinfo{person}{Roberto
  Perdisci}, {and} \bibinfo{person}{Manos Antonakakis}.}
  \bibinfo{year}{2015}\natexlab{}.
\newblock \showarticletitle{Segugio: Efficient Behavior-Based Tracking of
  Malware-Control Domains in Large {ISP} Networks}. In
  \bibinfo{booktitle}{\emph{2015 45th Annual IEEE/IFIP Int’l. Conf. on
  Dependable Systems and Networks}}. \bibinfo{publisher}{IEEE},
  \bibinfo{pages}{403--414}.
\newblock


\bibitem[Rezende et~al\mbox{.}(2014)]%
        {Rezende2014}
\bibfield{author}{\bibinfo{person}{Danilo~Jimenez Rezende},
  \bibinfo{person}{Shakir Mohamed}, {and} \bibinfo{person}{Daan Wierstra}.}
  \bibinfo{year}{2014}\natexlab{}.
\newblock \showarticletitle{Stochastic Backpropagation and Approximate
  Inference in Deep Generative Models}. In
  \bibinfo{booktitle}{\emph{Proceedings of the 31st International Conference on
  International Conference on Machine Learning - Volume 32}} (Beijing, China)
  \emph{(\bibinfo{series}{ICML'14})}. \bibinfo{publisher}{JMLR.org},
  \bibinfo{pages}{II–1278–II–1286}.
\newblock


\bibitem[Ribeiro et~al\mbox{.}(2015)]%
        {Ribeiro2015MLaaS}
\bibfield{author}{\bibinfo{person}{Mauro Ribeiro}, \bibinfo{person}{Katarina
  Grolinger}, {and} \bibinfo{person}{Miriam~A.M. Capretz}.}
  \bibinfo{year}{2015}\natexlab{}.
\newblock \showarticletitle{{MLaaS}: Machine Learning as a Service}. In
  \bibinfo{booktitle}{\emph{2015 IEEE 14th International Conference on Machine
  Learning and Applications (ICMLA)}}. \bibinfo{pages}{896--902}.
\newblock
\urldef\tempurl%
\url{https://doi.org/10.1109/ICMLA.2015.152}
\showDOI{\tempurl}


\bibitem[Ribeiro et~al\mbox{.}(2016)]%
        {ribeiro2016should}
\bibfield{author}{\bibinfo{person}{Marco~Tulio Ribeiro},
  \bibinfo{person}{Sameer Singh}, {and} \bibinfo{person}{Carlos Guestrin}.}
  \bibinfo{year}{2016}\natexlab{}.
\newblock \showarticletitle{" Why should i trust you?" Explaining the
  predictions of any classifier}. In \bibinfo{booktitle}{\emph{Proceedings of
  the 22nd ACM SIGKDD international conference on knowledge discovery and data
  mining}}. \bibinfo{pages}{1135--1144}.
\newblock


\bibitem[Severi et~al\mbox{.}(2021)]%
        {severiExplanationGuidedBackdoorPoisoning2021}
\bibfield{author}{\bibinfo{person}{Giorgio Severi}, \bibinfo{person}{Jim
  Meyer}, \bibinfo{person}{Scott Coull}, {and} \bibinfo{person}{Alina Oprea}.}
  \bibinfo{year}{2021}\natexlab{}.
\newblock \showarticletitle{Explanation-{{Guided Backdoor Poisoning Attacks
  Against Malware Classifiers}}}. In \bibinfo{booktitle}{\emph{30th {{USENIX
  Security Symposium}} ({{USENIX Security}} 21)}}. \bibinfo{pages}{1487--1504}.
\newblock
\showISBNx{978-1-939133-24-3}


\bibitem[Shafahi et~al\mbox{.}(2018)]%
        {shafahiPoisonFrogsTargeted2018}
\bibfield{author}{\bibinfo{person}{Ali Shafahi}, \bibinfo{person}{W.~Ronny
  Huang}, \bibinfo{person}{Mahyar Najibi}, \bibinfo{person}{Octavian Suciu},
  \bibinfo{person}{Christoph Studer}, \bibinfo{person}{Tudor Dumitras}, {and}
  \bibinfo{person}{Tom Goldstein}.} \bibinfo{year}{2018}\natexlab{}.
\newblock \showarticletitle{Poison {{Frogs}}! {{Targeted Clean-Label Poisoning
  Attacks}} on {{Neural Networks}}}. In \bibinfo{booktitle}{\emph{Advances in
  {{Neural Information Processing Systems}}}}.
\newblock


\bibitem[Sharafaldin et~al\mbox{.}(2018)]%
        {sharafaldinGeneratingNewIntrusion2018}
\bibfield{author}{\bibinfo{person}{Iman Sharafaldin}, \bibinfo{person}{Arash
  Habibi~Lashkari}, {and} \bibinfo{person}{Ali~A. Ghorbani}.}
  \bibinfo{year}{2018}\natexlab{}.
\newblock \showarticletitle{Toward {{Generating}} a {{New Intrusion Detection
  Dataset}} and {{Intrusion Traffic Characterization}}:}. In
  \bibinfo{booktitle}{\emph{Proceedings of the 4th {{International Conference}}
  on {{Information Systems Security}} and {{Privacy}}}}.
  \bibinfo{publisher}{{SCITEPRESS - Science and Technology Publications}},
  \bibinfo{address}{{Funchal, Madeira, Portugal}}, \bibinfo{pages}{108--116}.
\newblock
\showISBNx{978-989-758-282-0}
\urldef\tempurl%
\url{https://doi.org/10.5220/0006639801080116}
\showDOI{\tempurl}


\bibitem[Sheatsley et~al\mbox{.}(2021)]%
        {Sheatsley2021DomainConstraints}
\bibfield{author}{\bibinfo{person}{Ryan Sheatsley}, \bibinfo{person}{Blaine
  Hoak}, \bibinfo{person}{Eric Pauley}, \bibinfo{person}{Yohan Beugin},
  \bibinfo{person}{Michael~J. Weisman}, {and} \bibinfo{person}{Patrick
  McDaniel}.} \bibinfo{year}{2021}\natexlab{}.
\newblock \showarticletitle{On the Robustness of Domain Constraints}. In
  \bibinfo{booktitle}{\emph{Proceedings of the 2021 ACM SIGSAC Conference on
  Computer and Communications Security}} (Virtual Event, Republic of Korea)
  \emph{(\bibinfo{series}{CCS '21})}. \bibinfo{publisher}{Association for
  Computing Machinery}, \bibinfo{address}{New York, NY, USA},
  \bibinfo{pages}{495–515}.
\newblock
\showISBNx{9781450384544}
\urldef\tempurl%
\url{https://doi.org/10.1145/3460120.3484570}
\showDOI{\tempurl}


\bibitem[Shrikumar et~al\mbox{.}(2017)]%
        {shrikumar2017learning}
\bibfield{author}{\bibinfo{person}{Avanti Shrikumar}, \bibinfo{person}{Peyton
  Greenside}, {and} \bibinfo{person}{Anshul Kundaje}.}
  \bibinfo{year}{2017}\natexlab{}.
\newblock \showarticletitle{Learning important features through propagating
  activation differences}. In \bibinfo{booktitle}{\emph{International
  conference on machine learning}}. PMLR, \bibinfo{pages}{3145--3153}.
\newblock


\bibitem[Siva~Kumar et~al\mbox{.}(2020)]%
        {sivakumarAdversarialMachineLearningIndustry2020}
\bibfield{author}{\bibinfo{person}{Ram~Shankar Siva~Kumar},
  \bibinfo{person}{Magnus Nystr{\"o}m}, \bibinfo{person}{John Lambert},
  \bibinfo{person}{Andrew Marshall}, \bibinfo{person}{Mario Goertzel},
  \bibinfo{person}{Andi Comissoneru}, \bibinfo{person}{Matt Swann}, {and}
  \bibinfo{person}{Sharon Xia}.} \bibinfo{year}{2020}\natexlab{}.
\newblock \showarticletitle{Adversarial {{Machine Learning-Industry
  Perspectives}}}. In \bibinfo{booktitle}{\emph{2020 {{IEEE Security}} and
  {{Privacy Workshops}} ({{SPW}})}}. \bibinfo{pages}{69--75}.
\newblock
\urldef\tempurl%
\url{https://doi.org/10.1109/SPW50608.2020.00028}
\showDOI{\tempurl}


\bibitem[Tamersoy et~al\mbox{.}(2014)]%
        {tamersoy14}
\bibfield{author}{\bibinfo{person}{Acar Tamersoy}, \bibinfo{person}{Kevin
  Roundy}, {and} \bibinfo{person}{Duen~Horng Chau}.}
  \bibinfo{year}{2014}\natexlab{}.
\newblock \showarticletitle{Guilt by Association: Large Scale Malware Detection
  by Mining File-Relation Graphs}. In \bibinfo{booktitle}{\emph{Proceedings of
  the 20th ACM SIGKDD International Conference on Knowledge Discovery and Data
  Mining}} (New York, New York, USA) \emph{(\bibinfo{series}{KDD '14})}.
  \bibinfo{publisher}{Association for Computing Machinery},
  \bibinfo{address}{New York, NY, USA}, \bibinfo{pages}{1524–1533}.
\newblock
\showISBNx{9781450329569}
\urldef\tempurl%
\url{https://doi.org/10.1145/2623330.2623342}
\showDOI{\tempurl}


\bibitem[Tran et~al\mbox{.}(2018)]%
        {tranSpectralSignaturesBackdoor2018}
\bibfield{author}{\bibinfo{person}{Brandon Tran}, \bibinfo{person}{Jerry Li},
  {and} \bibinfo{person}{Aleksander M{\k{a}}dry}.}
  \bibinfo{year}{2018}\natexlab{}.
\newblock \showarticletitle{Spectral Signatures in Backdoor Attacks}. In
  \bibinfo{booktitle}{\emph{Proceedings of the 32nd {{International
  Conference}} on {{Neural Information Processing Systems}}}}
  \emph{(\bibinfo{series}{{{NIPS}}'18})}. \bibinfo{publisher}{{Curran
  Associates Inc.}}, \bibinfo{address}{{Montr\'eal, Canada}},
  \bibinfo{pages}{8011--8021}.
\newblock


\bibitem[Turner et~al\mbox{.}(2018)]%
        {turnerCleanLabelBackdoorAttacks2018}
\bibfield{author}{\bibinfo{person}{Alexander Turner}, \bibinfo{person}{Dimitris
  Tsipras}, {and} \bibinfo{person}{Aleksander M{\k{a}}dry}.}
  \bibinfo{year}{2018}\natexlab{}.
\newblock \showarticletitle{Clean-{{Label Backdoor Attacks}}}.
\newblock \bibinfo{journal}{\emph{Manuscript submitted for publication}}
  (\bibinfo{year}{2018}), \bibinfo{pages}{21}.
\newblock


\bibitem[Turner et~al\mbox{.}(2019)]%
        {turner2019labelconsistent}
\bibfield{author}{\bibinfo{person}{Alexander Turner}, \bibinfo{person}{Dimitris
  Tsipras}, {and} \bibinfo{person}{Aleksander Madry}.}
  \bibinfo{year}{2019}\natexlab{}.
\newblock \bibinfo{title}{Label-Consistent Backdoor Attacks}.
\newblock
\newblock
\showeprint[arxiv]{1912.02771}~[stat.ML]


\bibitem[Vargas~Muñoz et~al\mbox{.}(2018)]%
        {Muñoz2018Bayesian}
\bibfield{author}{\bibinfo{person}{María Vargas~Muñoz},
  \bibinfo{person}{Rafael Martínez-Peláez}, \bibinfo{person}{Pablo
  Velarde~Alvarado}, \bibinfo{person}{Efraín Moreno-Garcia},
  \bibinfo{person}{Deni Torres-Roman}, {and} \bibinfo{person}{José
  Ceballos-Mejia}.} \bibinfo{year}{2018}\natexlab{}.
\newblock \showarticletitle{Classification of network anomalies in flow level
  network traffic using Bayesian networks}. In \bibinfo{booktitle}{\emph{2018
  International Conference on Electronics, Communications and Computers
  (CONIELECOMP)}}. \bibinfo{publisher}{IEEE}, \bibinfo{pages}{238--243}.
\newblock
\urldef\tempurl%
\url{https://doi.org/10.1109/CONIELECOMP.2018.8327205}
\showDOI{\tempurl}


\bibitem[Wu et~al\mbox{.}(2019)]%
        {Wu2019Evading}
\bibfield{author}{\bibinfo{person}{Di Wu}, \bibinfo{person}{Binxing Fang},
  \bibinfo{person}{Junnan Wang}, \bibinfo{person}{Qixu Liu}, {and}
  \bibinfo{person}{Xiang Cui}.} \bibinfo{year}{2019}\natexlab{}.
\newblock \showarticletitle{Evading Machine Learning Botnet Detection Models
  via Deep Reinforcement Learning}. In \bibinfo{booktitle}{\emph{ICC 2019 -
  2019 IEEE International Conference on Communications (ICC)}}.
  \bibinfo{publisher}{IEEE}, \bibinfo{address}{Shanghai, China},
  \bibinfo{pages}{1--6}.
\newblock
\urldef\tempurl%
\url{https://doi.org/10.1109/ICC.2019.8761337}
\showDOI{\tempurl}


\bibitem[Xu and Shelton(2010)]%
        {Xu2010BayesianIDS}
\bibfield{author}{\bibinfo{person}{Jing Xu} {and} \bibinfo{person}{Christian~R.
  Shelton}.} \bibinfo{year}{2010}\natexlab{}.
\newblock \showarticletitle{Intrusion Detection Using Continuous Time Bayesian
  Networks}.
\newblock \bibinfo{journal}{\emph{J. Artif. Int. Res.}} \bibinfo{volume}{39},
  \bibinfo{number}{1} (\bibinfo{date}{sep} \bibinfo{year}{2010}),
  \bibinfo{pages}{745–774}.
\newblock
\showISSN{1076-9757}


\bibitem[Xu et~al\mbox{.}(2019a)]%
        {CTGAN_Xu2019}
\bibfield{author}{\bibinfo{person}{Lei Xu}, \bibinfo{person}{Maria
  Skoularidou}, \bibinfo{person}{Alfredo Cuesta-Infante}, {and}
  \bibinfo{person}{Kalyan Veeramachaneni}.} \bibinfo{year}{2019}\natexlab{a}.
\newblock \showarticletitle{{Modeling Tabular Data Using Conditional GAN}}. In
  \bibinfo{booktitle}{\emph{Proceedings of the 33rd International Conference on
  Neural Information Processing Systems}}. \bibinfo{publisher}{Curran
  Associates Inc.}, \bibinfo{address}{Red Hook, NY, USA}, Article
  \bibinfo{articleno}{659}, \bibinfo{numpages}{11}~pages.
\newblock


\bibitem[Xu et~al\mbox{.}(2019b)]%
        {xuModelingTabularData2019}
\bibfield{author}{\bibinfo{person}{Lei Xu}, \bibinfo{person}{Maria
  Skoularidou}, \bibinfo{person}{Alfredo {Cuesta-Infante}}, {and}
  \bibinfo{person}{Kalyan Veeramachaneni}.} \bibinfo{year}{2019}\natexlab{b}.
\newblock \showarticletitle{Modeling {{Tabular}} Data Using {{Conditional
  GAN}}}. In \bibinfo{booktitle}{\emph{Advances in {{Neural Information
  Processing Systems}}}}, Vol.~\bibinfo{volume}{32}.
  \bibinfo{publisher}{{Curran Associates, Inc.}}
\newblock


\bibitem[Yang et~al\mbox{.}(2021)]%
        {yangFeatureExtractionNovelty2021}
\bibfield{author}{\bibinfo{person}{Kun Yang}, \bibinfo{person}{Samory Kpotufe},
  {and} \bibinfo{person}{Nick Feamster}.} \bibinfo{year}{2021}\natexlab{}.
\newblock \bibinfo{title}{Feature {{Extraction}} for {{Novelty Detection}} in
  {{Network Traffic}}}.
\newblock
\newblock
\urldef\tempurl%
\url{https://doi.org/10.48550/arXiv.2006.16993}
\showDOI{\tempurl}
\showeprint[arxiv]{2006.16993}~[cs]


\bibitem[Yang et~al\mbox{.}(2023)]%
        {yangJigsawPuzzleSelective2023}
\bibfield{author}{\bibinfo{person}{Limin Yang}, \bibinfo{person}{Zhi Chen},
  \bibinfo{person}{Jacopo Cortellazzi}, \bibinfo{person}{Feargus Pendlebury},
  \bibinfo{person}{Kevin Tu}, \bibinfo{person}{Fabio Pierazzi},
  \bibinfo{person}{Lorenzo Cavallaro}, {and} \bibinfo{person}{Gang Wang}.}
  \bibinfo{year}{2023}\natexlab{}.
\newblock \showarticletitle{Jigsaw {{Puzzle}}: {{Selective Backdoor Attack}} to
  {{Subvert Malware Classifiers}}}. In \bibinfo{booktitle}{\emph{{{IEEE
  Symposium}} on {{Security}} \& {{Privacy}}}}.
\newblock


\bibitem[Young et~al\mbox{.}(2009)]%
        {Young2009Bayesian}
\bibfield{author}{\bibinfo{person}{Jim Young}, \bibinfo{person}{Patrick
  Graham}, {and} \bibinfo{person}{Richard Penny}.}
  \bibinfo{year}{2009}\natexlab{}.
\newblock \showarticletitle{Using Bayesian Networks to Create Synthetic Data}.
\newblock \bibinfo{journal}{\emph{Journal of Official Statistics}}
  \bibinfo{volume}{25} (\bibinfo{date}{12} \bibinfo{year}{2009}),
  \bibinfo{pages}{549--567}.
\newblock


\bibitem[Zhao et~al\mbox{.}(2021)]%
        {CTAB-GAN-zhao21a}
\bibfield{author}{\bibinfo{person}{Zilong Zhao}, \bibinfo{person}{Aditya
  Kunar}, \bibinfo{person}{Robert Birke}, {and} \bibinfo{person}{Lydia~Y.
  Chen}.} \bibinfo{year}{2021}\natexlab{}.
\newblock \showarticletitle{{CTAB-GAN}: Effective Table Data Synthesizing}. In
  \bibinfo{booktitle}{\emph{Proceedings of The 13th Asian Conference on Machine
  Learning}} \emph{(\bibinfo{series}{Proceedings of Machine Learning Research},
  Vol.~\bibinfo{volume}{157})}, \bibfield{editor}{\bibinfo{person}{Vineeth~N.
  Balasubramanian} {and} \bibinfo{person}{Ivor Tsang}} (Eds.).
  \bibinfo{publisher}{PMLR}, \bibinfo{pages}{97--112}.
\newblock
\urldef\tempurl%
\url{https://proceedings.mlr.press/v157/zhao21a.html}
\showURL{%
\tempurl}


\end{thebibliography}

\end{document}